\documentclass[namedreferences]{SolarPhysics}
\usepackage[optionalrh]{spr-sola-addons}
\usepackage{graphicx}
\usepackage{url}             
\usepackage{color}

\usepackage{solaheader}
\newcommand{\solareceiveddate}{14 July 2009}
\newcommand{\solaaccepteddate}{14 September 2009}
\newcommand{\solaonlinedate}{27 October 2009}
\newcommand{\doi}{009-9454-2}
\renewcommand{\solayear}{2009}
\renewcommand{\TODAY}{3 November 2009}

\newcommand{\grad}{\mbox{\boldmath$\nabla$}}
\newcommand{\vdot}{{\mathbf{\cdot}}}
\newcommand{\B}{\mathbf{B}}

\newcommand{\bxi}{B_x^{\rm i}}
\newcommand{\byi}{B_y^{\rm i}}
\newcommand{\bzi}{B_z^{\rm i}}
\newcommand{\bxh}{B_x^{\rm h}}
\newcommand{\byh}{B_y^{\rm h}}
\newcommand{\bzh}{B_z^{\rm h}}
\newcommand{\bhoriz}{B_{\rm h}}
\newcommand{\xuh}{x^{\rm h}}
\newcommand{\yuh}{y^{\rm h}}
\newcommand{\zuh}{z^{\rm h}}
\newcommand{\xui}{x^{\rm i}}
\newcommand{\yui}{y^{\rm i}}
\newcommand{\zui}{z^{\rm i}}
\newcommand{\myie}{{\it i.e., }}
\newcommand{\myeg}{{\it e.g., }}
\newcommand{\myetal}{{\it et al.}}
\begin{document}
\begin{article}
\begin{opening}

\title{Resolving the Azimuthal Ambiguity in Vector Magnetogram Data with the Divergence-Free Condition: Application to Discrete Data}
\author{A.D.~\surname{Crouch}\sep G.~\surname{Barnes}\sep K.D.~\surname{Leka}}

\runningauthor{A.D.~Crouch \myetal}
\runningtitle{Resolving the $180^\circ$ Ambiguity in Vector Magnetograms}

\institute{A.D.~Crouch\sep G.~Barnes\sep K.D.~Leka\\
NorthWest Research Associates, Colorado Research Associates Division, 3380 Mitchell Lane, Boulder, CO 80301, USA\\A.D.~Crouch\\email: \url{ash@cora.nwra.com}\\G.~Barnes\\email: \url{graham@cora.nwra.com}\\K.D.~Leka\\email: \url{leka@cora.nwra.com}}


\begin{abstract}
We investigate how the divergence-free property of magnetic fields can be exploited to resolve the azimuthal ambiguity  present in solar vector magnetogram data, by using line-of-sight and horizontal heliographic derivative information as approximated from discrete measurements.
Using synthetic data we test several methods that each make different assumptions about how the divergence-free property can be used to resolve the ambiguity.
We find that the most robust algorithm involves the minimisation of the absolute value of the divergence summed over the entire field of view.
Away from disk centre this method requires the sign and magnitude of the line-of-sight derivatives of all three components of the magnetic field vector.
\end{abstract}


\end{opening}

\section{Introduction}

Measurements of the vector magnetic field in the solar atmosphere are crucial for understanding a wide variety of physical phenomena.
Using the linear polarisation of magnetically sensitive spectral lines to infer the component of the field transverse to the line-of-sight results in an ambiguity of 180$^\circ$ in its direction \cite{1969PhDT3H}.
This ambiguity must be resolved in order to completely determine the magnetic field vector.
Several methods are currently in use to resolve the ambiguity (for overviews see \opencite{metcalfetal06}; \opencite{lekaetal2008}).
In general, each method involves an assumption or approximation that may not be appropriate for observations of solar magnetic fields.

A promising approach, which can potentially avoid unrealistic assumptions, is to exploit the divergence-free property of magnetic fields (\myeg \opencite{wuai1990}; \opencite{1993AA...278..279C}; \opencite{1993AA...279..214L}; \opencite{1999AA...347.1005B}; \opencite{lietal07}; \opencite{cb2007}).
One challenge in calculating the divergence is that it requires information about the variation of the field in the direction perpendicular to the solar surface.
To this end it is possible to derive from observations the variation of the field along the  line-of-sight direction
(\myeg \opencite{ruizcobodeltooroiniesta92}; \opencite{1994AA...291..622C}; \opencite{1995ApJ...439..474M}; \opencite{1996SoPh..164..169D}; \opencite{1996SoPh..169...79L}; \opencite{1998ApJ...494..453W}, \citeyear{2001ApJ...547.1130W}; \opencite{2000ApJ...530..977S}; \opencite{eibeetal2002}; \opencite{lekametcalf2003}; \opencite{2005ApJ...631L.167S}, \citeyear{2007ApJS..169..439S}).
The line-of-sight direction is not perpendicular to the solar surface, except at disk centre; therefore some care is required to correctly treat the geometrical perspective.

In \citeauthor{cb2007}~(\citeyear{cb2007}, henceforth paper~I) we showed that
the  ambiguity  can be resolved with
line-of-sight and horizontal heliographic derivative information
by using the divergence-free property  without additional assumptions about the nature of the field.
To resolve the ambiguity away from disk centre we demonstrated that it is sufficient to determine
four pieces of information:
{\it i)}   the sign of the line-of-sight derivative of the line-of-sight component of the magnetic field;
{\it ii)}  the sign and magnitude of the line-of-sight derivative of the magnitude of the transverse component of the field;
{\it iii)} the sign and magnitude of the line-of-sight derivative of the azimuthal angle; and
{\it iv)}  the sign and magnitude of the horizontal heliographic derivatives of the magnitude of the transverse component of the field and the azimuthal angle.
Paper~I was a theoretical examination in which we assumed that all  necessary derivatives could be determined exactly at a given location and that the solar surface could be represented by the heliographic plane.

The aim of this paper is to investigate how the divergence-free property can be used to resolve the ambiguity for data where the magnetic field is measured at discrete spatial locations.
Using synthetic data we test two methods that have previously appeared in the literature (\myeg \opencite{wuai1990}; \opencite{1993AA...278..279C}; \opencite{1993AA...279..214L}; \opencite{1999AA...347.1005B}; \opencite{lietal07}); both of these are based on the divergence but they  make different assumptions about how to use it to resolve the ambiguity.
We show that both of these methods can produce significant errors.
For this reason, we present a more robust algorithm, the global minimisation method.

The outline of this paper is as follows.
In Section~\ref{sec_divb} we describe the derivation of the divergence-free condition for any position on the solar disk in terms of observable quantities and discuss the consequences for ambiguity resolution.
In Section~\ref{sec_synth} we review the synthetic data and metrics that will be used to test the performance of the ambiguity resolution algorithms.
In Section~\ref{algorithms} we test the algorithms and the validity of the assumptions made.
In Section~\ref{sec_conc} we draw conclusions.

\section{The Divergence-Free Condition}
\label{sec_divb}

Within a limited field of view we assume that the solar surface can be represented by the heliographic plane, which is tangent to the solar surface at the position given by the central meridian angle, the latitude, and the $P$- and $B_0$-angles.
Over the heliographic plane, the image components of the magnetic field are related to the heliographic components according to the following transformation:

\begin{eqnarray}
\bxh & = & a_{11} \bxi + a_{12} \byi + a_{13} \bzi \, , \nonumber \\
\byh & = & a_{21} \bxi + a_{22} \byi + a_{23} \bzi \, , \label{B_h} \\
\bzh & = & a_{31} \bxi + a_{32} \byi + a_{33} \bzi \, ,  \nonumber
\end{eqnarray}

\noindent
where the coefficients $a_{ij}$ are taken from Equation~(1) of \inlinecite{1990SoPh..126...21G}, $\bzi = B_\|$ is the line-of-sight component of the field, and the components of the field perpendicular to the line-of-sight direction are given by

\begin{equation}
\bxi = B_\perp \cos \xi \, , \qquad \mbox{and} \qquad \byi = B_\perp \sin \xi \, ,
\label{btrans}
\end{equation}

\noindent
where $B_\perp$ is the magnitude of the component transverse to the line-of-sight and the azimuthal angle, $\xi$, is the angle between the transverse component of the field and the $\xui$-axis (where $\xui$ and $\yui$ are coordinates on the plane perpendicular to the line-of-sight). The azimuthal angle can only be determined observationally within the range $0 \leq \xi < 180^\circ$.

The derivative in the direction perpendicular to the heliographic plane, $\partial / \partial \zuh$,  cannot be directly measured away from disk centre. 
However, the derivative along the line-of-sight direction, $\partial / \partial \zui$, can be measured (see below for further discussion).
These derivatives are related as follows:

\begin{equation}
\frac{\partial f}{\partial \zui} = a_{13} \frac{\partial f}{ \partial \xuh} + a_{23} \frac{\partial f}{ \partial \yuh} + a_{33} \frac{\partial f}{\partial \zuh} \, ,
\label{dlos}
\end{equation}

\noindent
where $f$ is any differentiable function, $\zui$ is the distance along the line-of-sight direction, and $\xuh$ and $\yuh$ are coordinates on the heliographic plane.
Derivatives with respect to horizontal heliographic distance, $\partial / \partial \xuh$ and $\partial / \partial \yuh$, can be approximated with techniques such as finite differences and finite elements using discrete measurements over  the heliographic plane. 
For convenience, we work with the line-of-sight distance $\zui$, although
in practice the field is likely to be inferred as a function of
optical depth; in which case a model atmosphere must be used to determine the correspondence between line-of-sight distance and optical depth (\myeg \opencite{1986ApJ...306..284M}; \opencite{1981ApJS...45..635V}; \opencite{1994AA...291..622C}; \opencite{2007ApJS..169..439S}).

Equations~(\ref{B_h}) and (\ref{dlos}) can be used to express the divergence of the field in terms of observable quantities (\myie derivatives of the image components of the field with respect to $\xuh$, $\yuh$, and $\zui$),

\begin{equation}
a_{33} \grad \vdot \B  = D_a + a_{33} \frac{\partial  \bzi}{\partial \zui} \, , 
\label{divb2}
\end{equation}

\noindent
where

\begin{eqnarray}
D_a & = & a_{31} \frac{\partial \bxi}{\partial \zui} + a_{32} \frac{\partial \byi}{\partial \zui}
+ \left( a_{11} a_{33} - a_{13} a_{31} \right) \frac{\partial \bxi}{\partial \xuh}
+ \left( a_{12} a_{33} - a_{13} a_{32} \right) \frac{\partial \byi}{\partial \xuh} \nonumber \\
& + & \left( a_{21} a_{33} - a_{23} a_{31} \right) \frac{\partial \bxi}{\partial \yuh}
+ \left( a_{22} a_{33} - a_{23} a_{32} \right) \frac{\partial \byi}{\partial \yuh} \, . \label{da}
\end{eqnarray}

\noindent
Equations~(\ref{divb2}) and (\ref{da}) show that the computation of the divergence requires the line-of-sight derivatives of all three components of the magnetic field vector, except at disk centre where the coefficients $a_{ij} = \delta_{ij}$ (see paper~I for further discussion).

Several techniques have appeared in the literature to measure the line-of-sight variation of the magnetic field (often but not always including the full magnetic field vector).  
We briefly discuss three that could be employed to evaluate the various terms in Equations~(\ref{divb2}) and (\ref{da}).
The first and possibly most straightforward is to measure the field at different heights using multiple spectral lines (\myeg \opencite{1996SoPh..169...79L}; \opencite{lekametcalf2003}).
The inferred height difference between measurements can be several hundred kilometres or more depending on line choice, although exact heights cannot be easily quantified due to the lines' often broad range of formation height.
A second approach makes use of a single (generally  chromospheric) line that forms over a range of heights.
Inversions performed at multiple, but limited wavelength ranges near (for example) the core and separately in the line wing can return the field strengths at different atmospheric heights (\myeg \opencite{1995ApJ...439..474M}; \opencite{eibeetal2002}).
Again, estimates of the field are obtained at discrete but broad heights which may be separated by tens or hundreds of kilometres.
A third approach recovers the line-of-sight stratification of the magnetic field during the inversion of the Stokes spectra through a formalism based on response functions (\myeg \opencite{ruizcobodeltooroiniesta92}; \opencite{1994AA...291..622C}; \opencite{1996SoPh..164..169D}; \opencite{1998ApJ...494..453W}, \citeyear{2001ApJ...547.1130W}; \opencite{2000ApJ...530..977S}; \opencite{2005ApJ...631L.167S}, \citeyear{2007ApJS..169..439S}).
The field is computed at discrete heights (inversion nodes) which may be separated by several hundred kilometres (\myeg  \opencite{socasnavarro2004});
interpolation can map the results onto a grid with points separated by as little as 5 kilometres.
For all of the aforementioned techniques, the field is effectively sampled discretely. Therefore, the  line-of-sight derivatives in Equations~(\ref{divb2}) and (\ref{da}) must be approximated.

It is important to understand that the inferred line-of-sight derivatives of the transverse components of the field, $\partial \bxi / \partial \zui$ and $\partial \byi / \partial \zui$, depend on the ambiguity resolution at all heights used to approximate line-of-sight derivatives.
Consequently, for ambiguity resolution algorithms based on these derivatives, vector magnetogram data should be disambiguated simultaneously at all relevant heights.

\section{Synthetic Data and Performance Metrics}
\label{sec_synth}

\inlinecite{metcalfetal06} compared several techniques for disambiguating single-height magnetogram data for which information about the variation of the field out of the heliographic plane was not available.
To test the performance of the various algorithms based on the divergence we use the same magnetic field configurations that were used by \inlinecite{metcalfetal06} for which the correct ambiguity resolution is known and magnetic field data are available at two heights.
Methods based on the divergence require magnetic field data from at least two heights.
All methods examined in this paper could be adapted for data with more than two heights; however, this can add a degree of complexity to the disambiguation process (depending on the algorithm) because all heights may need to be disambiguated simultaneously.
For each synthetic magnetogram, the field is sampled at discrete spatial locations.
The grid points are intended to mimic pixel centres, although the exact value of the field is provided.
Thus, the effects of noise and unresolved structure in both the line-of-sight and horizontal heliographic directions are neglected (see \myeg \opencite{lekaetal2008}); these issues are beyond the scope of this investigation and will be considered in a separate paper.
The grid points at each height are aligned in the line-of-sight direction and separated by a line-of-sight distance, $\Delta \zui$.

\begin{figure}[ht]
\centerline{
\includegraphics[width=0.49\textwidth,viewport=10 0 470 320]{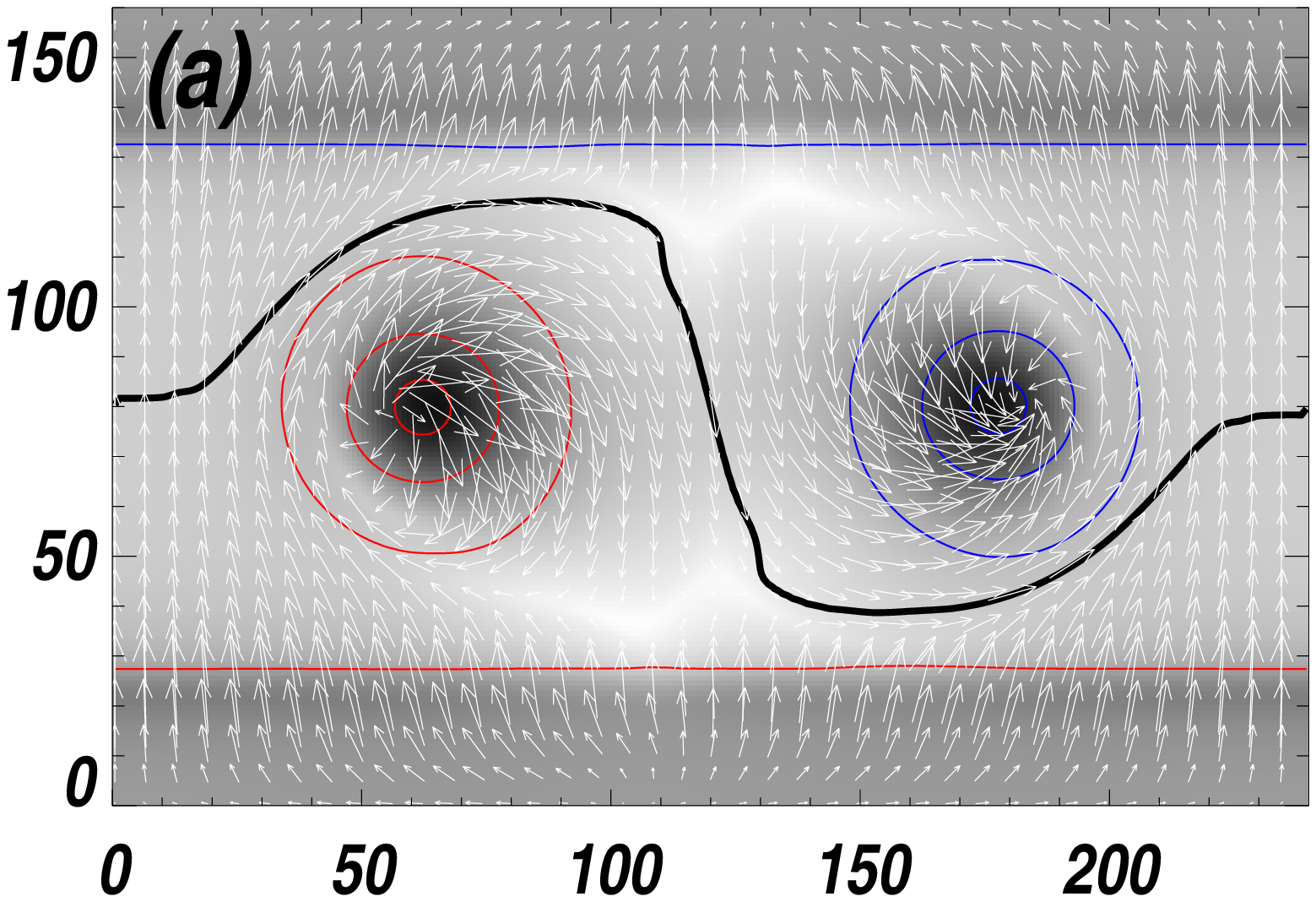}
\hspace*{0.02\textwidth}
\includegraphics[width=0.49\textwidth,viewport=10 15 465 465]{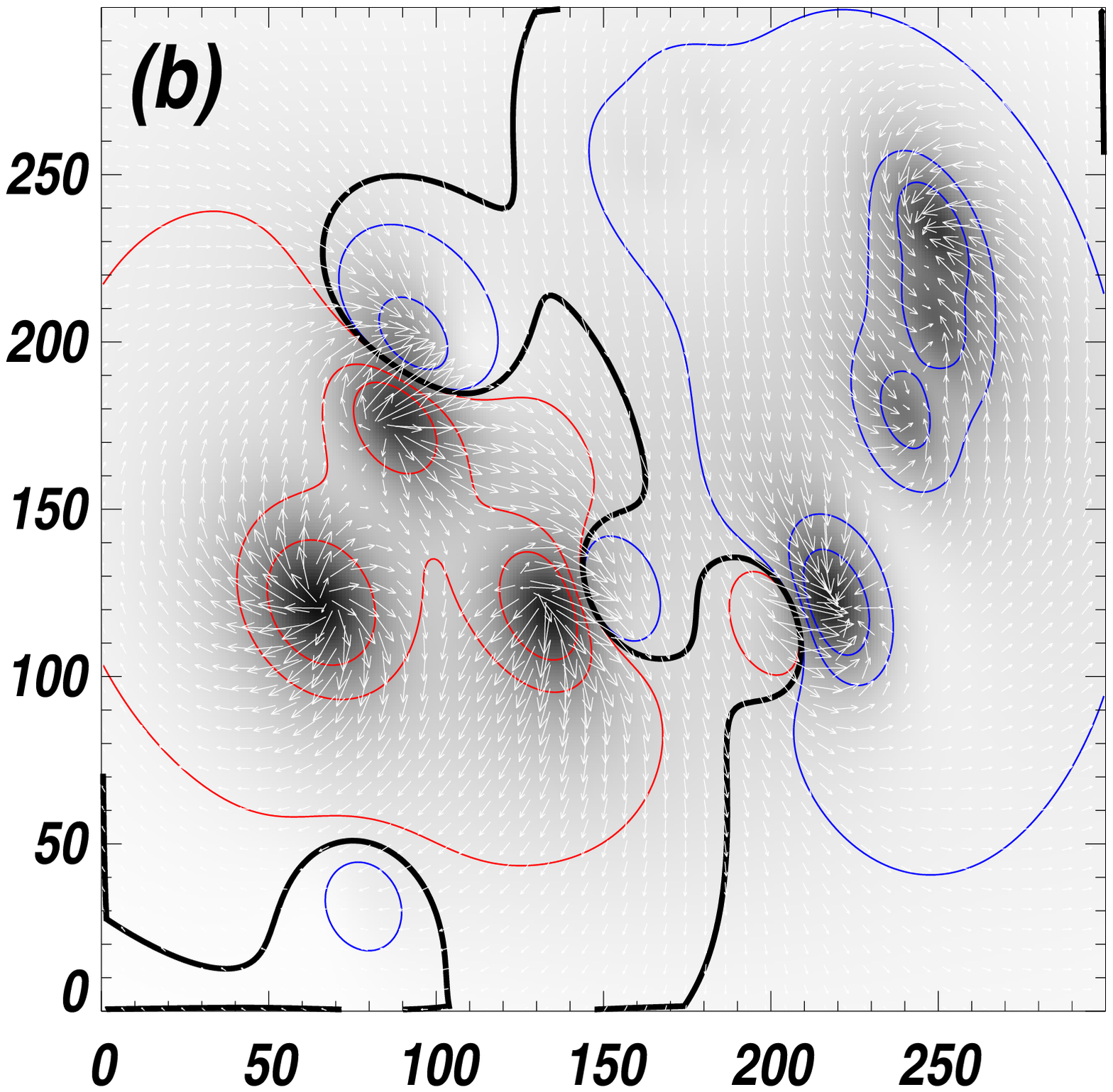}
}
\caption{Synthetic vector magnetic field data. 
In both panels the underlying ``continuum image'' is a reverse-color image of $B^2$;
positive/negative vertical magnetic flux is indicated by {\it red}/{\it blue} contours at 100, 1000, 2000~G, 
and the magnetic neutral line is indicated by the {\it black} contour.
Horizontal magnetic field is plotted at every fifth pixel, with magnitude proportional to arrow length. 
Tickmarks are in units of pixels.
(a) Vector magnetic field at $z = 0.00625 L$ from the numerical simulation of Fan and Gibson~(2004)
at their time step 56 ($L$ is a length scale, effectively the size of the simulation domain in the $y$-direction).
The pixel size is $0.00625 L$.
This is an exact reproduction of Figure~1 from Metcalf~\myetal~(2006).
(b) Vector magnetic field for the multipole structure in image coordinates.
The pixel size is $0.5''$.
This is an exact reproduction of Figure~2 from Metcalf~\myetal~(2006).
}
\label{synth}
\end{figure}

The field shown in Figure~\ref{synth}(a) is a snapshot from the numerical MHD simulation of \inlinecite{2004ApJ...609.1123F} of a highly twisted flux tube emerging into an overlying potential field ``arcade''. 
Because the simulation results are provided on a three-dimensional rectangular grid this case corresponds to disk centre, where the line-of-sight and $\zuh$-directions are parallel.
Figure~\ref{synth}(a) shows the field at the first height (as used by \opencite{metcalfetal06}); the second height used in this investigation lies above this, and is discussed in detail below.

The field displayed in Figure~\ref{synth}(b) is constructed from a collection of point sources located on a plane below the surface.
The contribution to the field from each source is calculated using the Green's function given by \inlinecite{1977ApJ...212..873C} with a constant value of the force-free parameter $\alpha$.
This magnetogram is located away from disk centre, centred at N $18^\circ$, W $45^\circ$, and it includes the effects of curvature.
Again Figure~\ref{synth}(b) shows the field at the first height (as used by \opencite{metcalfetal06}); the second height lies above this and is discussed further below.

To quantify the performance of the various algorithms we will use the same four metrics presented in \inlinecite{metcalfetal06}:
the fraction of pixels correctly disambiguated $\mathcal{M}_{\rm area}$, 
the fraction of magnetic flux with the correct solution $\mathcal{M}_{\rm flux}$, 
the fraction of strong horizontal field regions ($\bhoriz > 500$~G) with the correct solution $\mathcal{M}_{\rm h}$, and 
a measure of the discrepancy in the vertical current between the retrieved and correct solution, $\mathcal{M}_{J_z}$;
note that $\mathcal{M}_{J_z}$ can have values outside the range $\left[ 0 , 1 \right]$ unlike the other three metrics.
Thus the results presented here can be directly compared to those obtained for the  algorithms examined by \inlinecite{metcalfetal06}.

\section{Testing the Algorithms}
\label{algorithms}

In this section we test several disambiguation methods all based on the divergence-free condition but which employ different assumptions and implementation details.
First, we examine two methods that have previously appeared in the literature: 
the \citeauthor{wuai1990} criterion (\myeg \opencite{wuai1990}; \opencite{1993AA...278..279C}; \opencite{1993AA...279..214L};  \opencite{lietal07})
and the  sequential minimisation method (\myeg \opencite{1999AA...347.1005B}).
We find that both of these methods can produce significant errors for both synthetic data sets.
Consequently, we present and test a third algorithm, the global minimisation method.

\subsection{The Wu and Ai Criterion}
\label{sec_wuai}

One non-iterative approach that was originally proposed by \inlinecite{wuai1990} and tested on synthetic data by \inlinecite{1993AA...278..279C}, \inlinecite{1993AA...279..214L}, and \inlinecite{lietal07}, involves expressing the divergence-free condition as an inequality.
For any position on the solar disk (except the limb) the divergence-free condition can be written as

\begin{equation}
a_{33} D_a \frac{\partial \bzi}{\partial \zui} = - \left( a_{33} \frac{\partial \bzi }{\partial \zui} \right)^2 \le 0 \, .
\label{divb1}
\end{equation}

\noindent
Assuming that the various derivatives can be determined reliably, given the sign of $\partial \bzi / \partial \zui$ the ambiguity can in principle be resolved by choosing the azimuthal angle such that $D_a$ satisfies Equation~(\ref{divb1}).
If $\partial \bzi / \partial \zui = 0$ or $D_a = 0$  then Equation~(\ref{divb1}) cannot be used to resolve the ambiguity.
Consequently, for discrete observations Equation~(\ref{divb1}) cannot be used if $| \partial \bzi / \partial \zui | \lesssim \delta ( \partial \bzi / \partial \zui )$ or if $| D_a | \lesssim \delta ( D_a )$, where $\delta$ is an estimate of the measurement error in the respective derivatives.
One advantage of using Equation~(\ref{divb1}) to resolve the ambiguity is that only the sign of $\partial \bzi / \partial \zui$ is required, not its magnitude; note, however, that the sign and magnitude of  $\partial \bxi / \partial \zui$ and  $\partial \byi / \partial \zui$ are required away from disk centre (see paper~I).
One limitation of using Equation~(\ref{divb1}) is that the inferred divergence of the field is not exactly zero (in general) when approximated from discrete observations, so the inequality is not exact (\opencite{1993AA...278..279C}).

\begin{figure}[ht]
\centerline{
\includegraphics[width=0.49\textwidth]{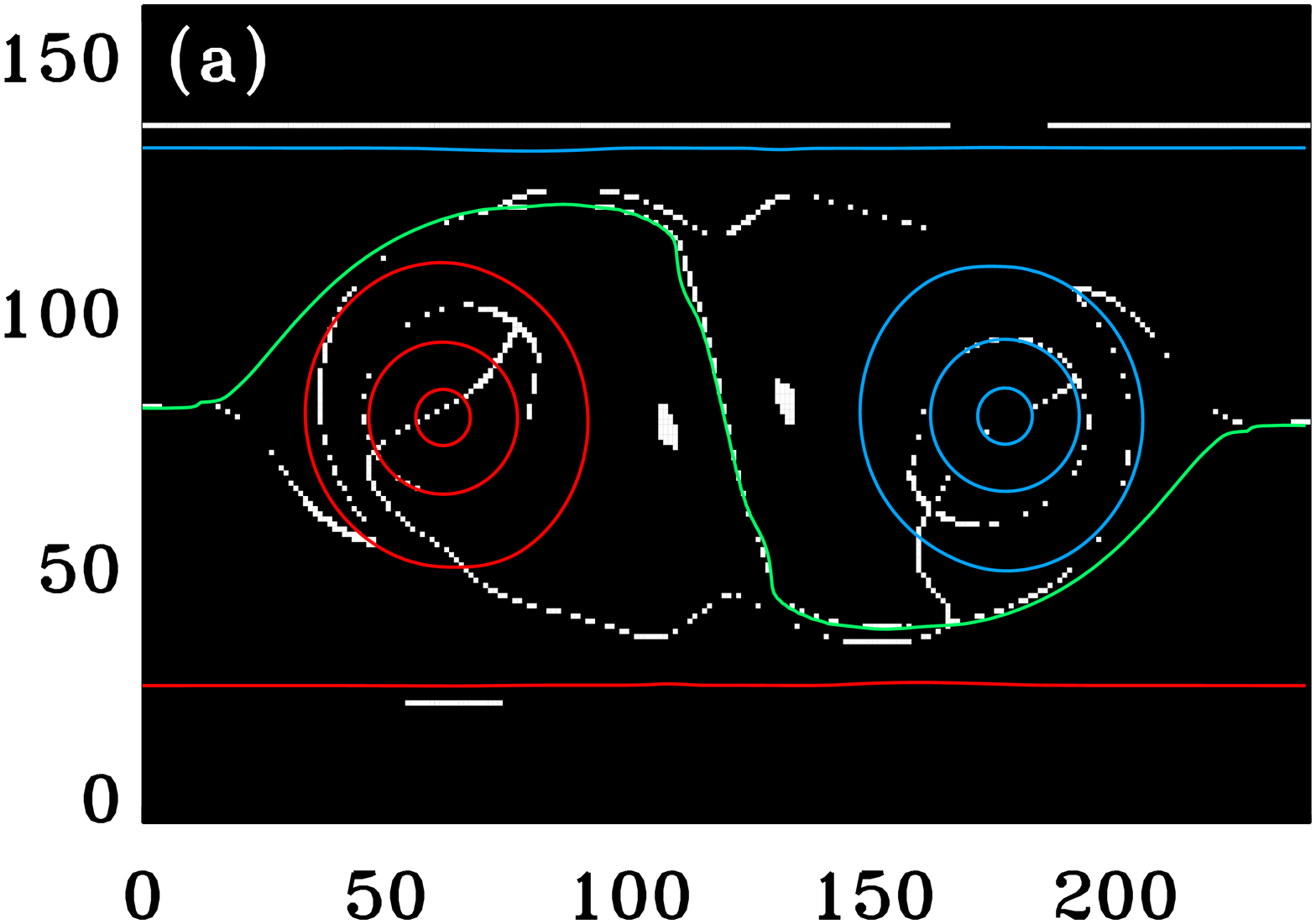}  
\hspace*{0.02\textwidth}
\includegraphics[width=0.49\textwidth]{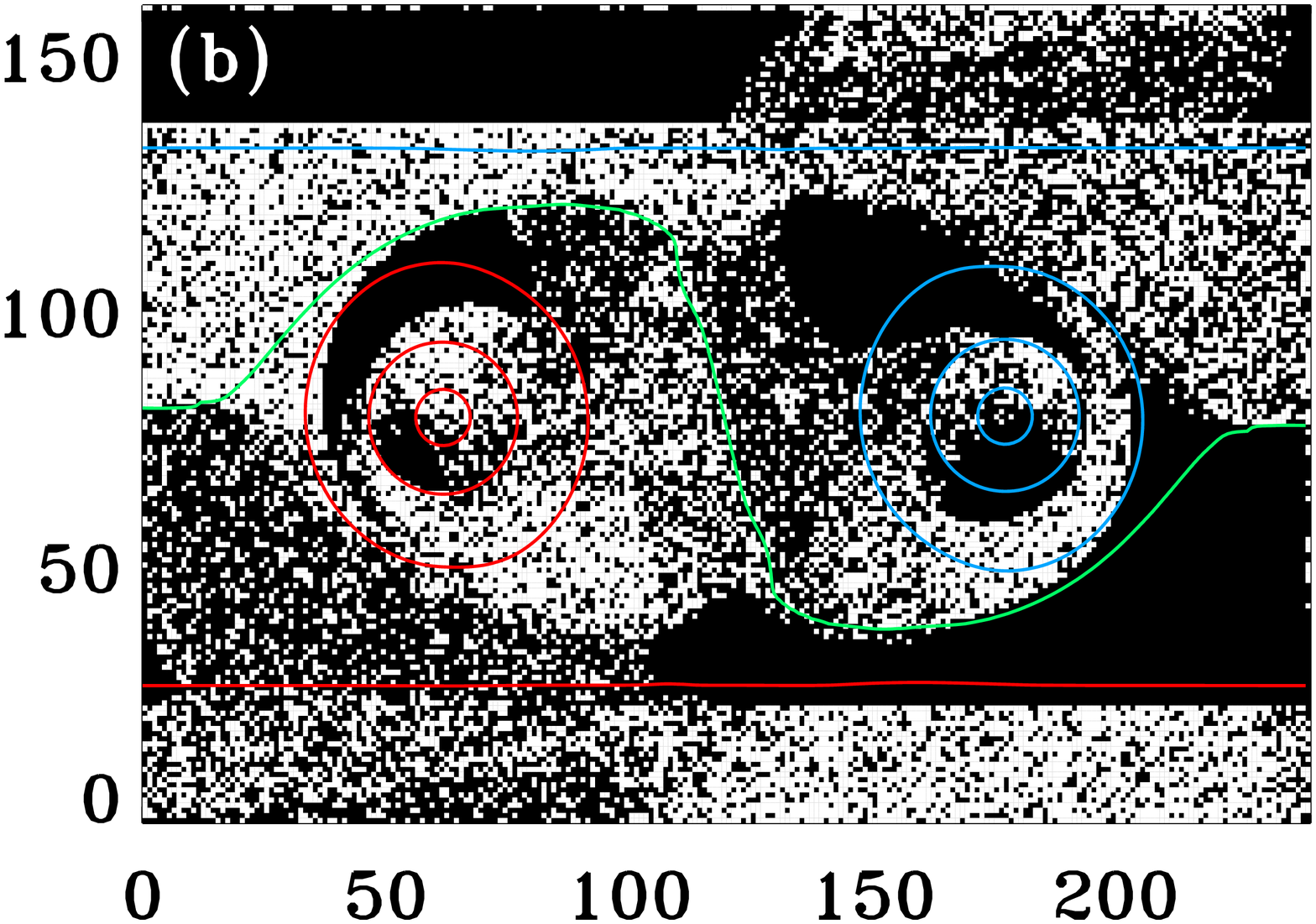}
}
\caption{(a) Disambiguation results for the Wu and Ai criterion applied to the flux tube and arcade configuration (Figure~\ref{synth}(a)).
{\it Contours} are the same as  Figure~\ref{synth}(a),  except that the {\it green} contour is the magnetic neutral line.
Areas with the correct/incorrect ambiguity resolution are {\it black}/{\it white}  (the same convention as Metcalf~\myetal, 2006).
Initial configuration has $\byi > 0$.
(b) Same as (a) except the initial configuration of azimuthal angles is random.}
\label{fan_ts56}
\end{figure}

The algorithm proceeds as follows.
At a given height each column of pixels is scanned from $j=1$ to $j=n_y$, and different columns are disambiguated from $i=1$  to $i=n_x$.
Hereafter, the indices $i$, $j$, and $k$ correspond to pixel labelling in the $\xui$-, $\yui$-, and $\zui$-directions, respectively, and $n_x$ and $n_y$ refer to the number of pixels in the $\xui$- and $\yui$-directions, respectively.
For $1 \leq i < n_x$ and $1 \leq j < n_y$ at a given height $k$ we approximate the horizontal heliographic derivatives $\partial / \partial \xuh$ and $\partial / \partial \yuh$ at pixel $(i,j,k)$ with three-point finite differences using the measurements at pixels: $(i,j,k)$, $(i+1,j,k)$, and $(i,j+1,k)$.
At disk centre, where the heliographic and image coordinates are equivalent, these approximations reduce to standard two-point forward finite differences.
At the boundaries $i=n_x$ and $j=n_y$ the geometry of the finite differencing stencil is modified to use only pixels within the field of view.
At the lower/upper height forward/backward finite differences are used to approximate the line-of-sight derivative $\partial / \partial \zui$ using measurements from the two heights.

\begin{table}[ht]
\caption{Performance metrics for the flux tube and arcade at disk centre for the various ambiguity resolution algorithms (with $\Delta \zui =1$~pixel).  The lower height is evaluated.}
\label{fan_tab}
\begin{tabular}{ccccc}
\hline
Method &  $\mathcal{M}_{\rm area}$ & $\mathcal{M}_{\rm flux}$ & $\mathcal{M}_{\rm h}$ &  $\mathcal{M}_{J_z}$ \\
\hline
\citeauthor{wuai1990}, initialised with $\byi > 0$                   & 0.98 & 0.98 & 0.97 & -0.29  \\ 
\citeauthor{wuai1990}, initialised by randomising $\xi$              & 0.62 & 0.63 & 0.63 & -15.24 \\ 
Sequential minimisation                                              & 0.94 & 0.93 & 0.93 & -0.88  \\ 
Global minimisation                                                  & 1.00 & 1.00 & 1.00 & 1.00 \\ 
\hline
\end{tabular}
\end{table}

Away from disk centre the lower height $k=1$ is examined first followed by the second height $k=2$.
At disk centre only the lower height is disambiguated; this is possible because $\partial \bxi / \partial \zui$ and $\partial \byi / \partial \zui$ are not required to compute the divergence at disk centre (see Equations~(\ref{divb2}) and (\ref{da})) and the approximations for horizontal heliographic derivatives involve measurements at a single height.
At each pixel $(i, j, k)$ we evaluate the quantities in Equation~(\ref{divb1}).
If the criterion is satisfied then no action is taken and we move to the next pixel in the sequence.
If the criterion is violated then the other choice of azimuthal angle is selected at pixel $(i, j, k)$.
To be consistent with the algorithms described in \inlinecite{wuai1990}, \inlinecite{1993AA...278..279C}, \inlinecite{1993AA...279..214L}, and \inlinecite{lietal07} we do not check that  $| \partial \bzi / \partial \zui |$ and $| D_a |$ exceed their respective measurement errors nor that Equation~(\ref{divb1}) is satisfied if the other choice of azimuthal angle is selected.

\begin{figure}[ht]
\centerline{
\includegraphics[width=0.49\textwidth]{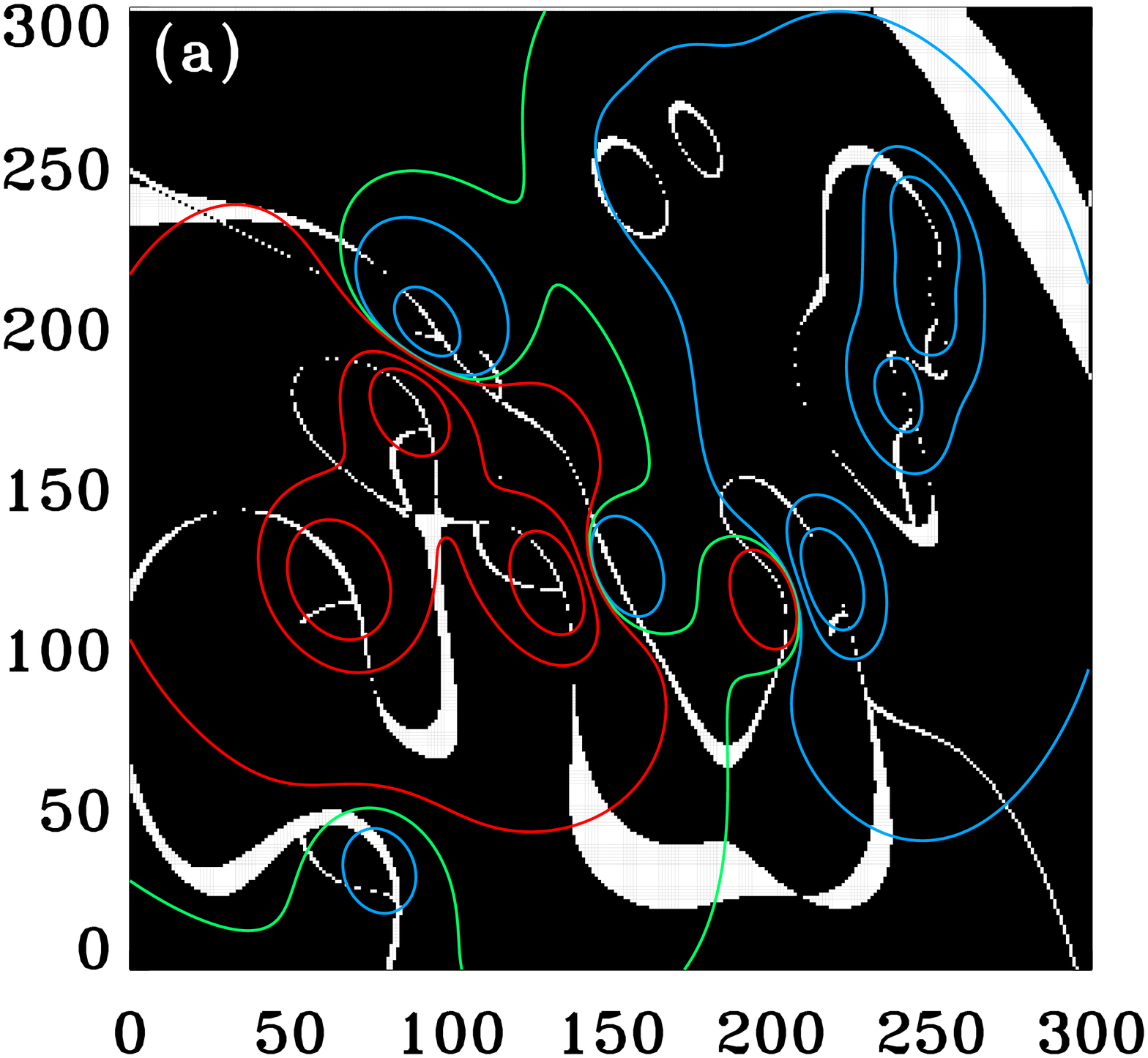}  
\hspace*{0.02\textwidth}
\includegraphics[width=0.49\textwidth]{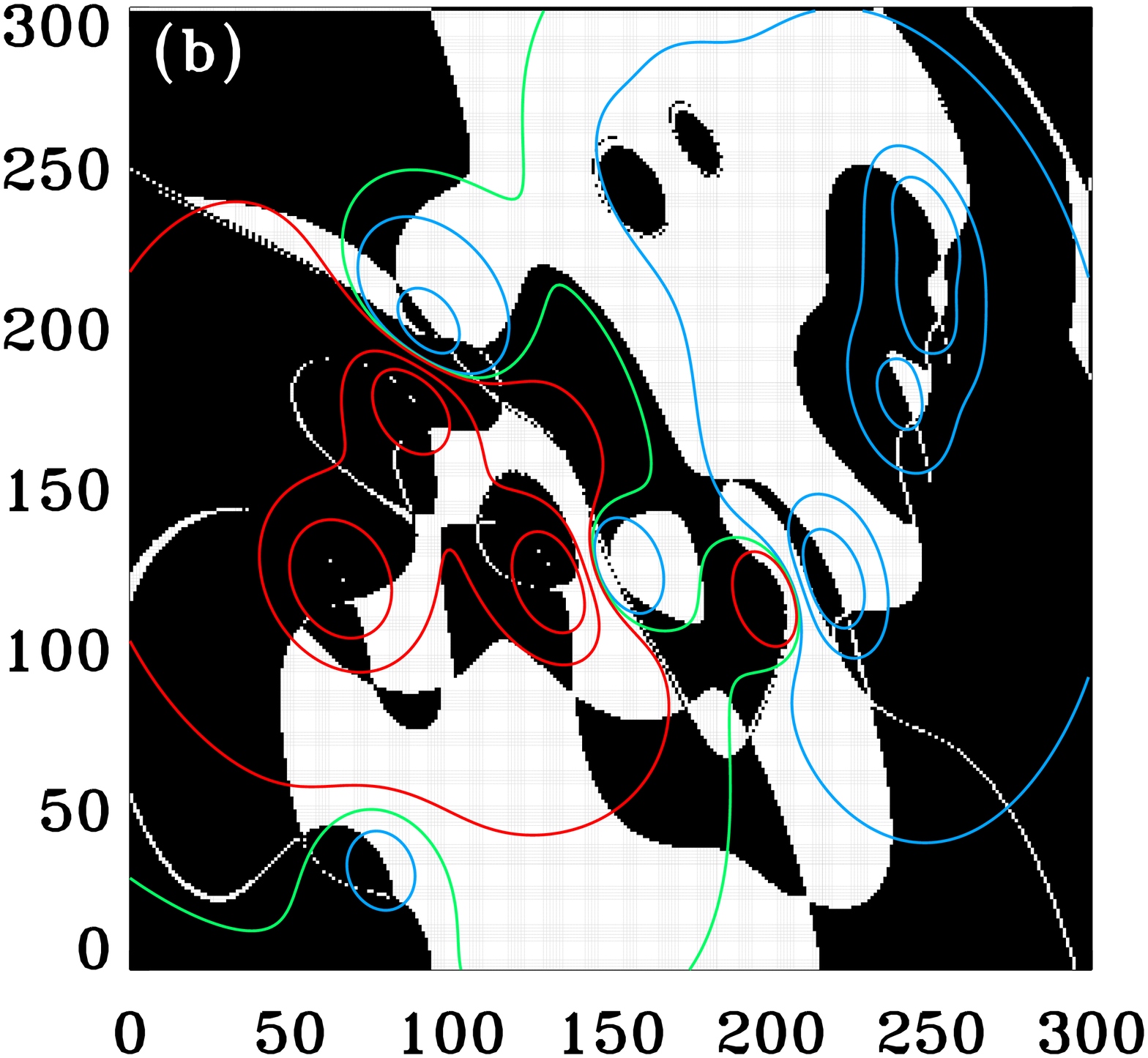}
}
\centerline{
\includegraphics[width=0.49\textwidth]{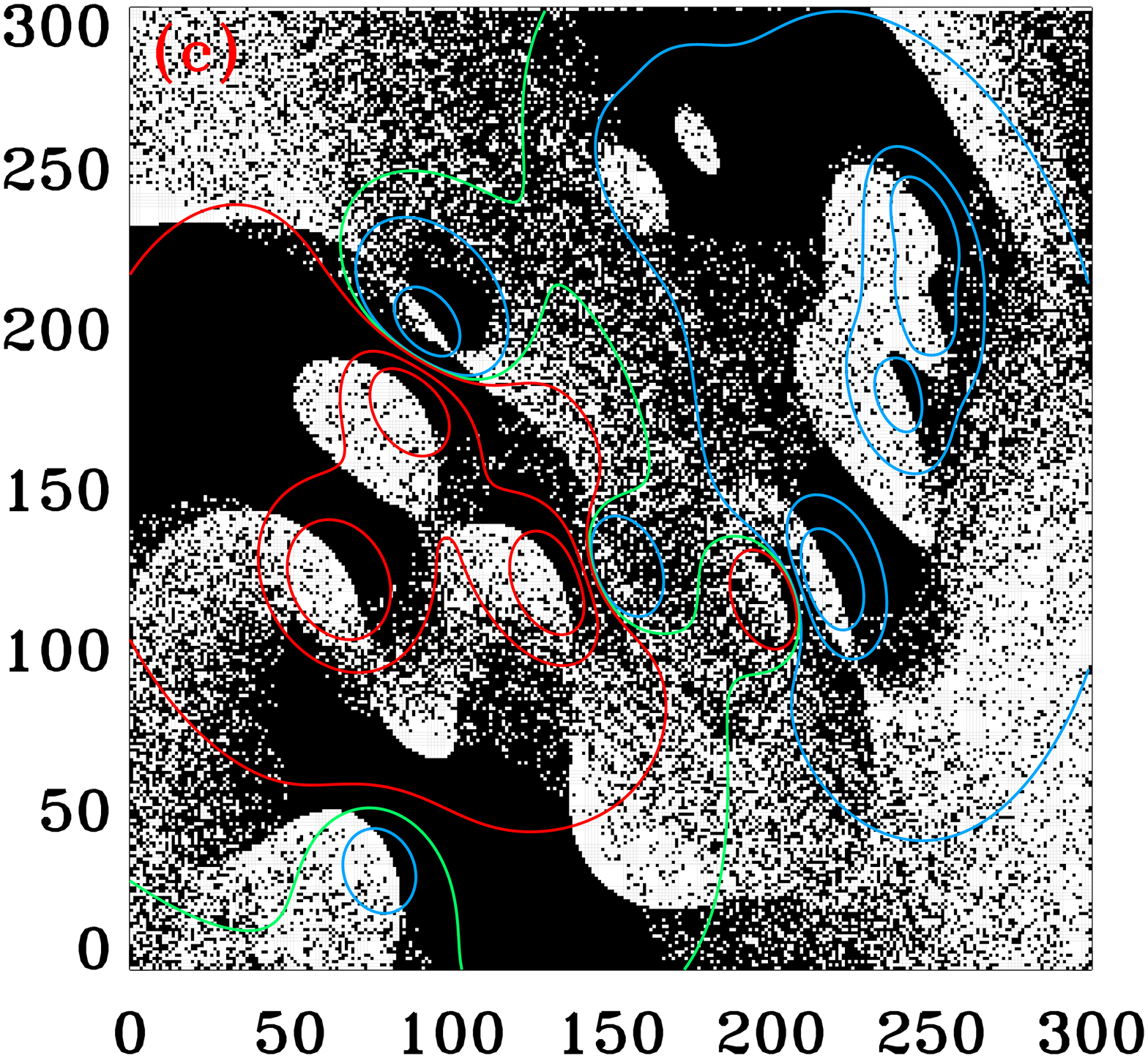}  
\hspace*{0.02\textwidth}
\includegraphics[width=0.49\textwidth]{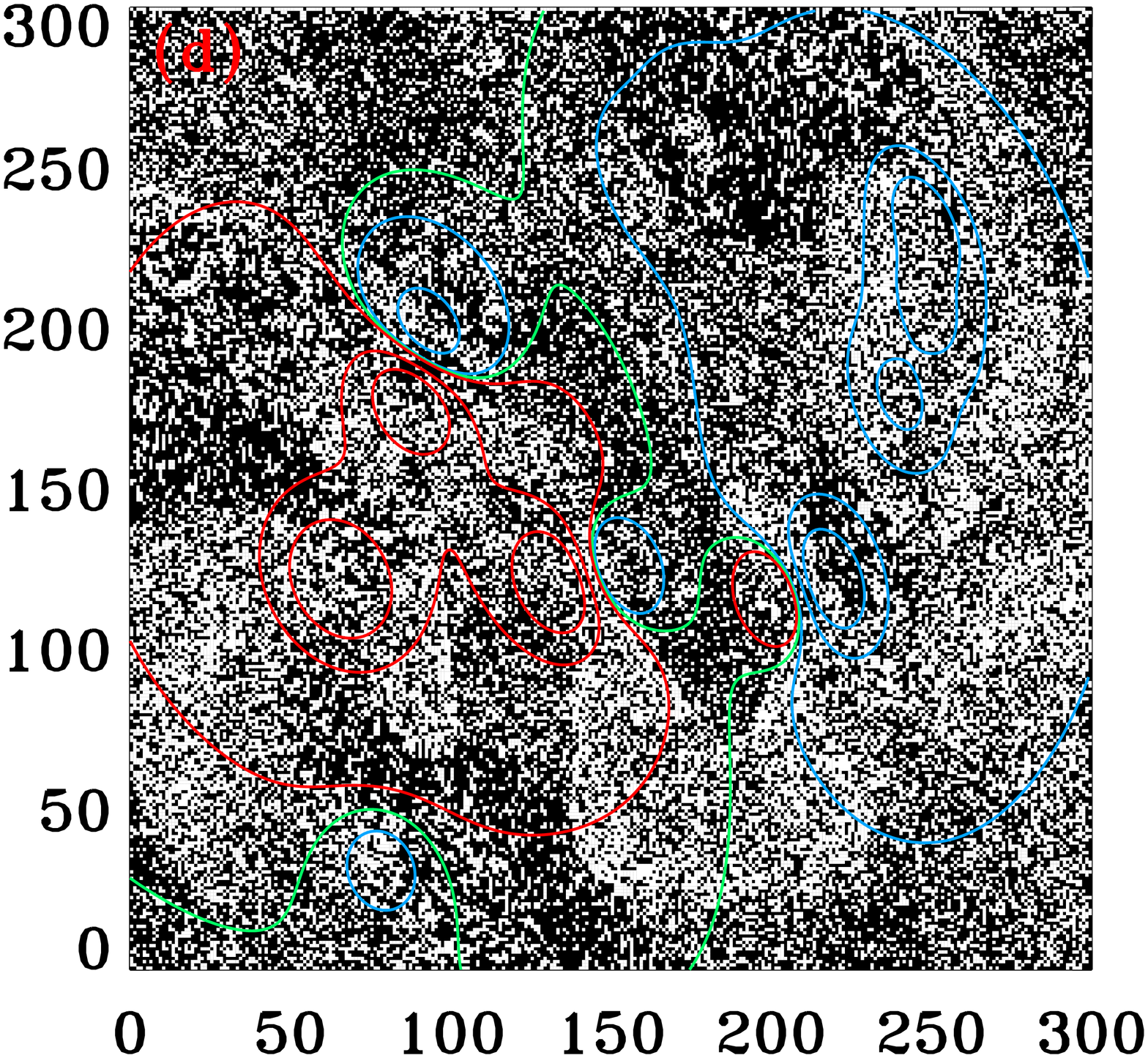}
}
\caption{(a) Disambiguation results for the Wu and Ai criterion applied to the multipole field configuration (Figure~\ref{synth}(b)), following  Figure~\ref{fan_ts56}(a).
(b) Same as (a) except for the second height, 1 pixel (or $0.5''$) above that shown in (a).
(c) Same as (a) except the initial configuration of azimuthal angles is random.
(d) Same as (c) except for the second height.
}
\label{compare_tpd7_wuai}
\end{figure}

To demonstrate the sensitivity of this algorithm to the initial configuration of azimuthal angles we show two cases.
First,  we initialise the data by setting $\byi >0$ at each pixel; this effectively smoothes the initial configuration.
At disk centre this is essentially the approach taken by \inlinecite{lietal07}, who also used synthetic data from the numerical simulation of \inlinecite{2004ApJ...609.1123F}, although there may be some subtle differences between our implementation and the \citeauthor{lietal07} implementation (such as the order in which pixels are examined and the exact heights used).
Second, we initialise the data by randomising the choice of azimuthal angle at each pixel (the initial choice of azimuthal angle is changed in approximately 50\% of pixels).
The performance of this algorithm for both initialisations is shown in Figures~\ref{fan_ts56} and \ref{compare_tpd7_wuai}, and Tables~\ref{fan_tab} and \ref{tpd7_tab}.

Evidently, this method is very sensitive to the smoothness of the initial configuration.
This is because the approximation for $D_a$ may be misleading when the initial configuration is not smooth.
This algorithm is also sensitive to the order in which pixels are examined: 
for the off-disk-centre data  (Figure~\ref{compare_tpd7_wuai}) the results at the upper height tend to be worse than those at the lower height.
This is because the smoothness of the initial configuration can be perturbed during the disambiguation process,  resulting in a misleading approximation for $D_a$.

\begin{table}[ht]
\caption{Performance metrics for the multipole field positioned away from disk centre for the various ambiguity resolution algorithms ($\Delta \zui =1$~pixel). The lower height is evaluated.}
\label{tpd7_tab}
\begin{tabular}{ccccc}
\hline
Method &  $\mathcal{M}_{\rm area}$ & $\mathcal{M}_{\rm flux}$ & $\mathcal{M}_{\rm h}$ &  $\mathcal{M}_{J_z}$ \\
\hline
\citeauthor{wuai1990}, initialised with $\byi > 0$                   & 0.92 & 0.94 & 0.93 & -1.14  \\ 
\citeauthor{wuai1990}, initialised by randomising $\xi$              & 0.61 & 0.59 & 0.66 & -13.70 \\ 
Sequential minimisation                                              & 0.95 & 0.93 & 0.94 & -3.36  \\ 
Global minimisation                                                  & 1.00 & 1.00 & 1.00 & 1.00   \\
\hline
\end{tabular}
\end{table}

\begin{figure}[ht]
\centerline{\includegraphics[width=0.49\textwidth]{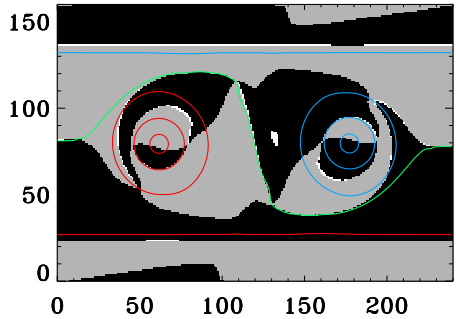}}
\caption{The Wu and Ai criterion applied to the correct solution for the flux tube and arcade configuration (Figure~\ref{synth}(a)).
{\it Contours} are the same as Figure~\ref{fan_ts56}.
Areas that satisfy/violate the Wu and Ai criterion, Equation~(\ref{divb1}), are coloured {\it black}/{\it white}.
Areas for which the sign of $D_a$ is the same for both choices of the azimuthal angle are coloured {\it gray}.}
\label{inequality}
\end{figure}

\begin{figure}[ht]
\centerline{
\includegraphics[width=0.49\textwidth]{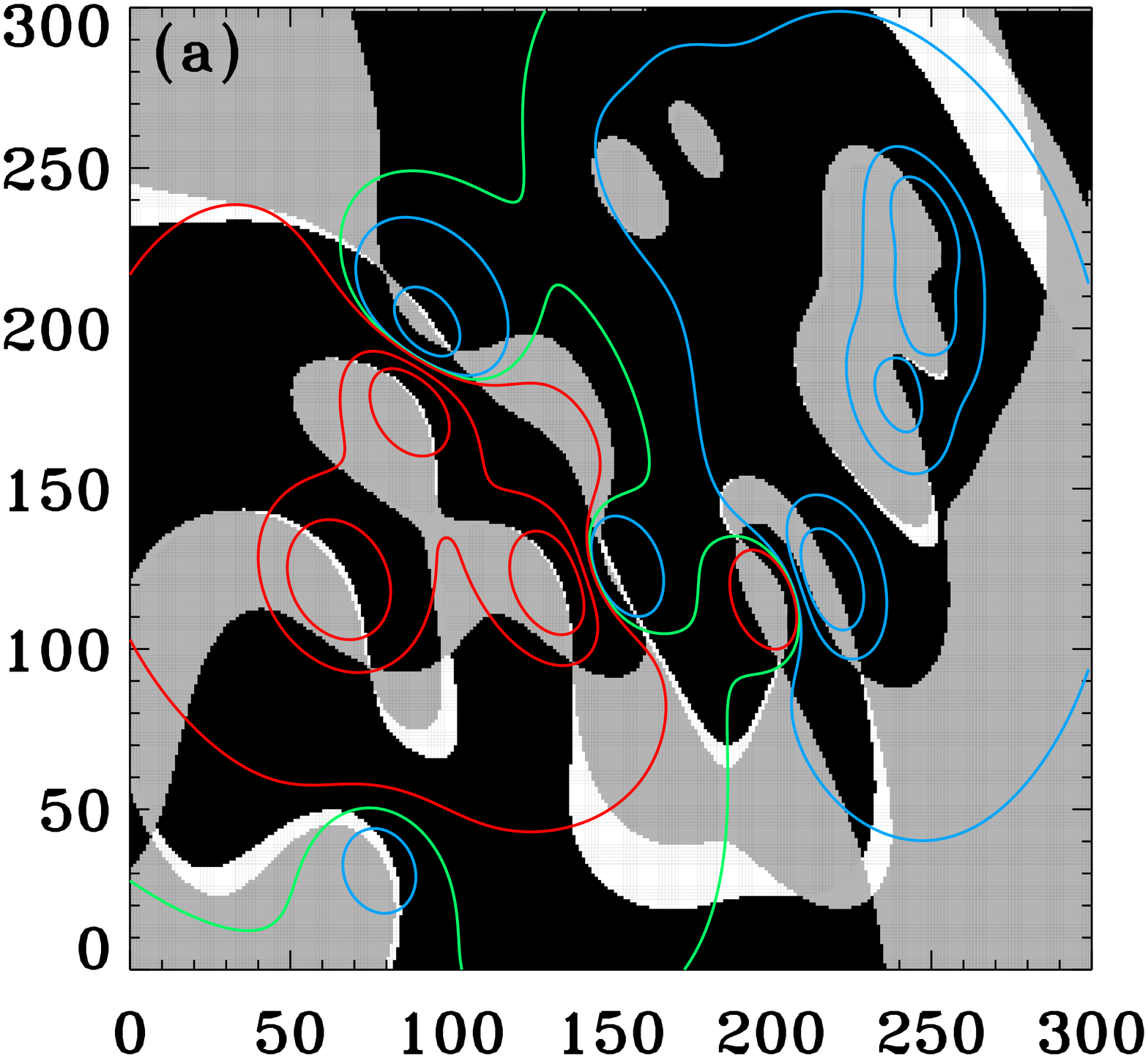}
\hspace*{0.02\textwidth}
\includegraphics[width=0.49\textwidth]{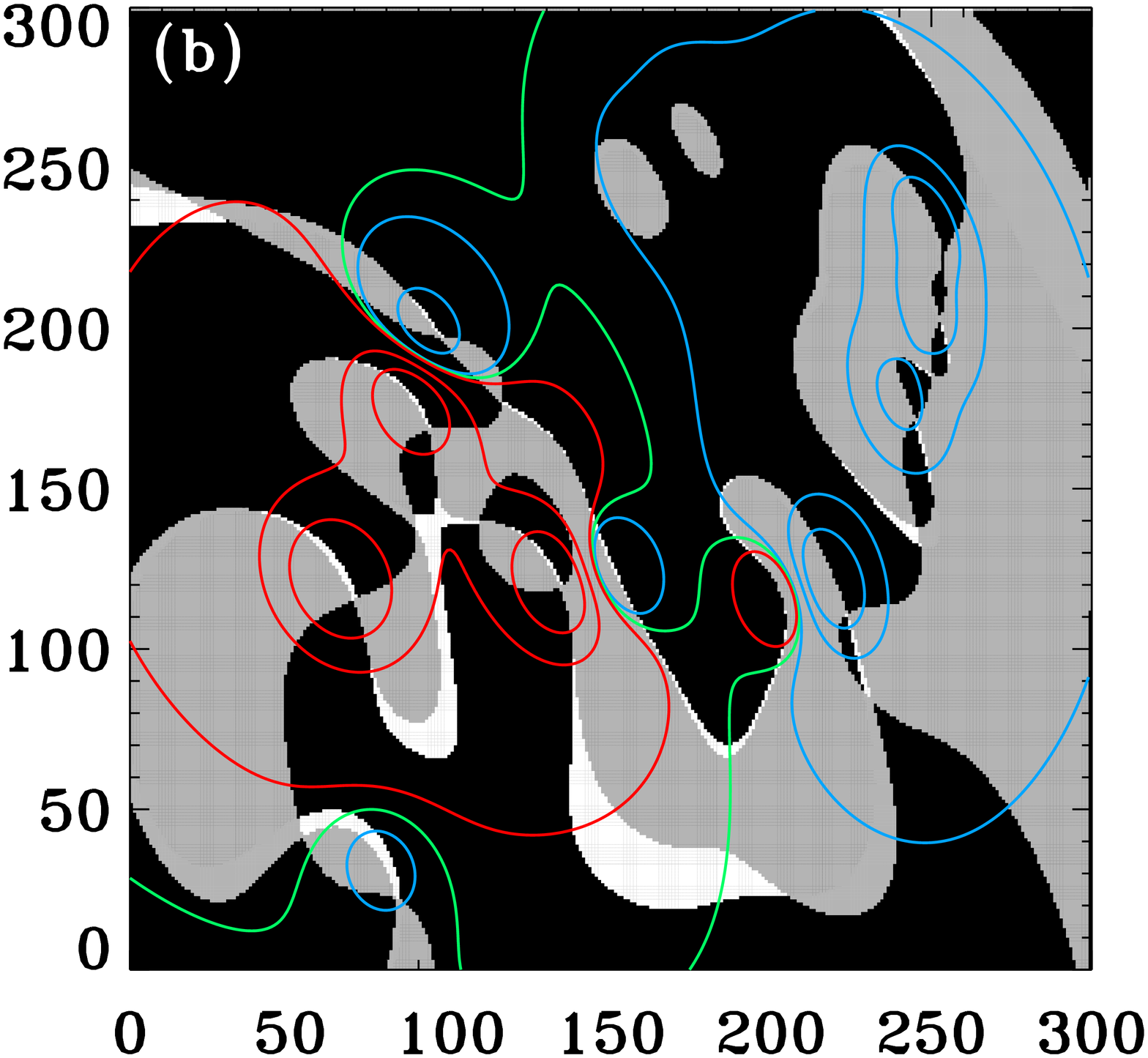}
}
\caption{(a) The Wu and Ai criterion applied to the correct solution for multipole field configuration at the lower height (Figure~\ref{synth}(b));  
{\it black}/{\it white}/{\it gray} indicate areas which satisfy, do not satisfy, and are degenerate to the inequality in Equation~(\ref{divb1}), following Figure~\ref{inequality}.
(b) Same as (a) except for the second height.
}
\label{inequality_tpd7}
\end{figure}

The assumption made by this method is tested by using the correct solution to evaluate each of the terms in Equation~(\ref{divb1}) with the various derivatives approximated as described above.
In Figures~\ref{inequality} and \ref{inequality_tpd7} areas that satisfy the inequality in Equation~(\ref{divb1}) for the correct solution are shown along with areas that violate Equation~(\ref{divb1}).
Also shown are areas which suffer from the degenerate case where Equation~(\ref{divb1}) cannot be used to distinguish the two realisations, as the sign of $D_a$ is the same for both the correct and the incorrect choice of azimuthal angle at the pixel of interest.
This is the case for a large fraction of the field of view for both synthetic data sets.
The method appears to do quite well in  Figure~\ref{fan_ts56}(a) and Figure~\ref{compare_tpd7_wuai}(a) because we do not test the other realisation if the initial one satisfies Equation~(\ref{divb1}) and we do not check that Equation~(\ref{divb1}) is satisfied after the choice of azimuthal angle is changed.
The degeneracy is partly due to the details of the implementation and could be avoided by changing the azimuthal angle in all pixels referenced by the finite differencing stencil at a particular location, rather than at a single pixel; unfortunately, such an approach would be very sensitive to the initial configuration.

\subsection{The Sequential Minimisation Method}

\inlinecite{1999AA...347.1005B} presented a method which assumes that the magnitude of the divergence is minimised over the set of pixels used to approximate the divergence at a particular location.
Disambiguation then proceeds as a sequence of minimisation problems based at each pixel in the field of view.

We apply this approach as follows.
At a given height, we examine pixels in the same order as in Section~\ref{sec_wuai}.
For the pixels referenced by the finite differencing stencil at a particular location we choose the permutation of azimuthal angles that corresponds to the minimum of $| ( \grad \vdot \B )_{i,j,k} |$, the approximation for the divergence at the pixel $(i, j, k)$.
We compute the divergence using the methods described in Section~\ref{sec_wuai}, with all approximations made about the lower height. 
The azimuthal angle at each pixel is fixed by the stencil that first references it.
This approach examines both permutations of the azimuthal angle at each pixel, therefore the initial configuration does not affect its performance, although it is sensitive to the order in which pixels are examined.

\begin{figure}[ht]
\centerline{
\includegraphics[width=0.49\textwidth]{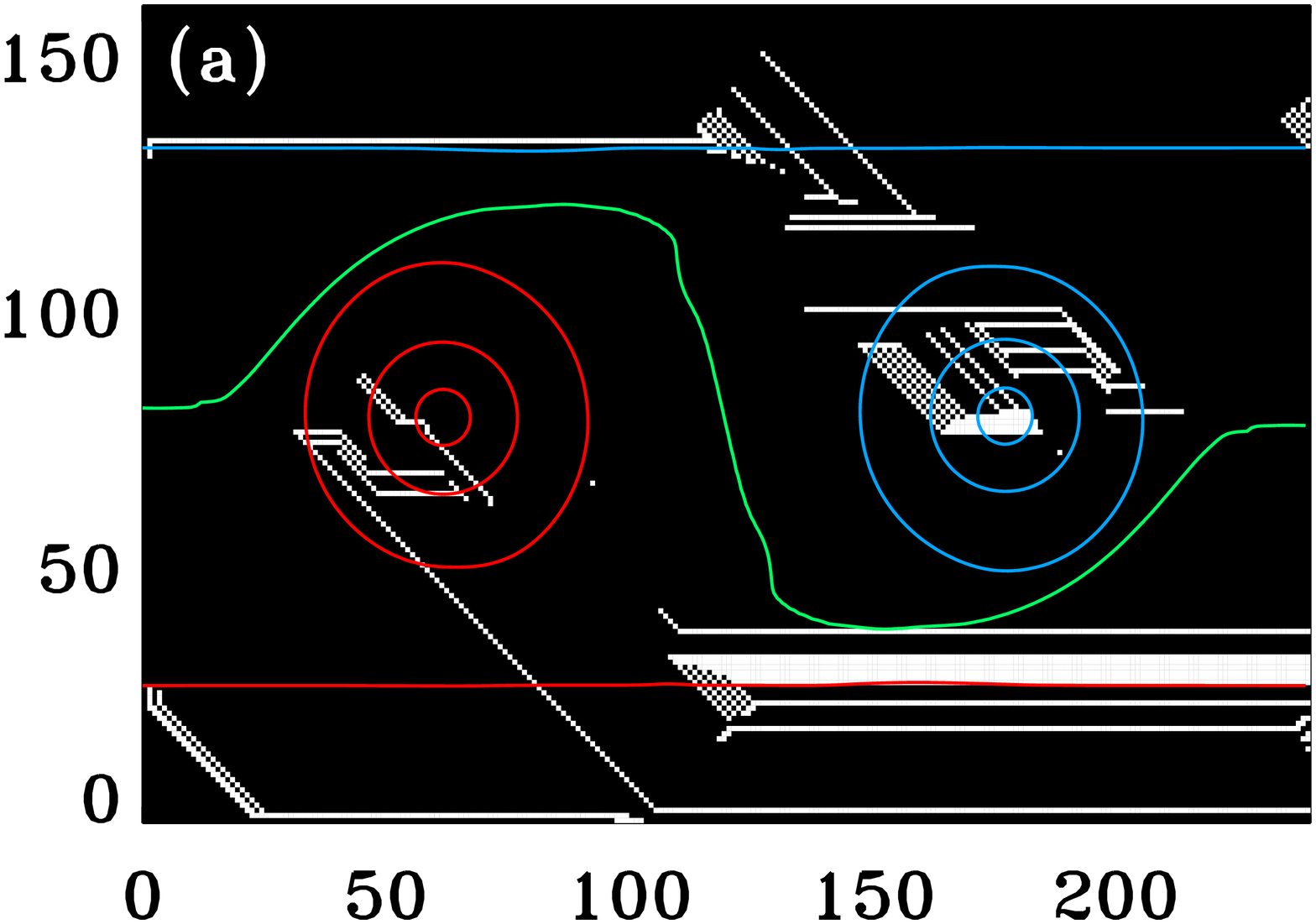}  
\hspace*{0.02\textwidth}
\includegraphics[width=0.49\textwidth]{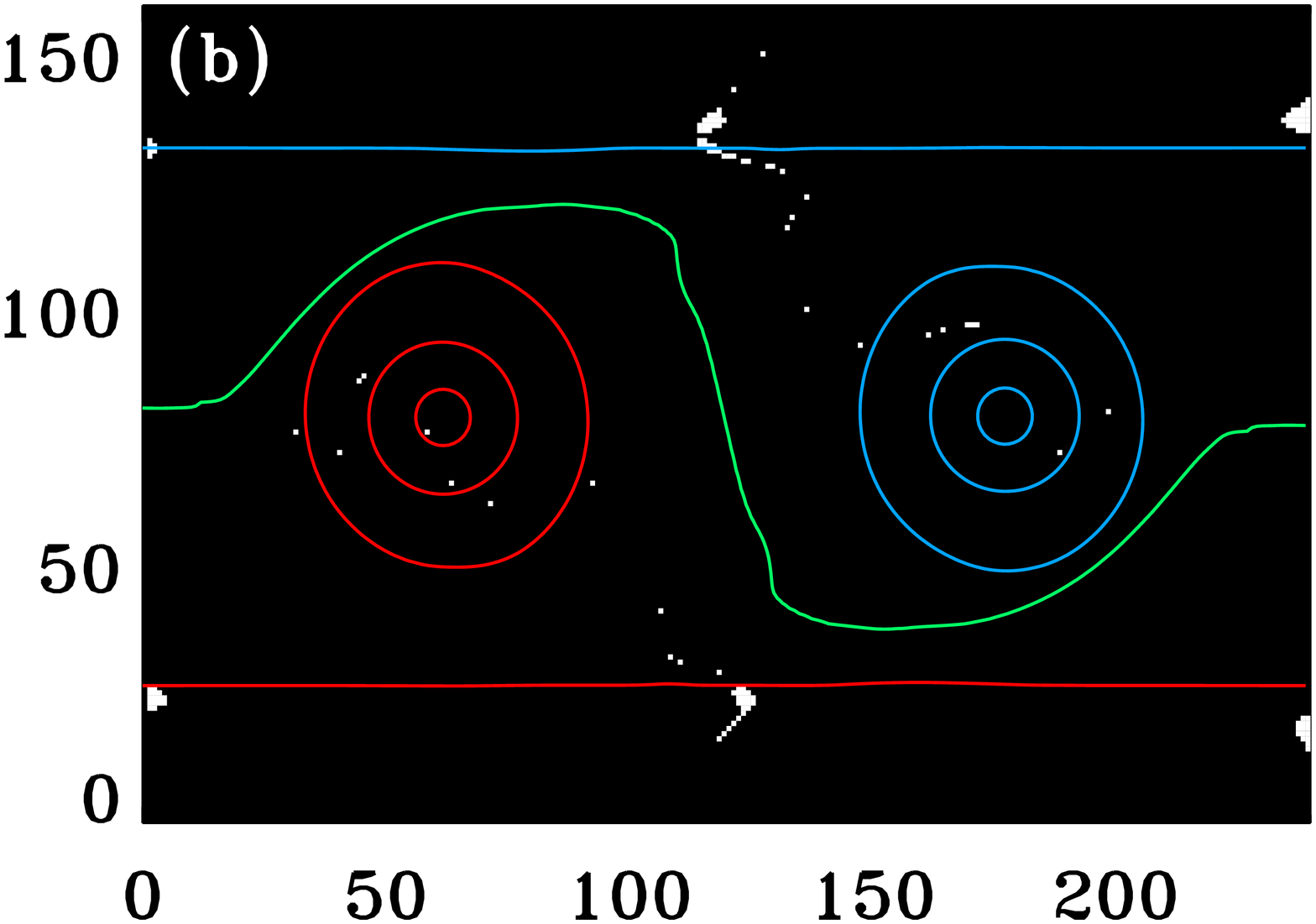}
}
\caption{(a) Disambiguation results for the sequential minimisation method applied to the flux tube and arcade configuration, following Figure~\ref{fan_ts56}.
(b) Validity of the assumption made by the sequential minimisation method.
Areas that satisfy/violate the assumption are coloured {\it black}/{\it white}.
{\it Contours} are the same as (a).
}
\label{compare_min_local}
\end{figure}

\begin{figure}[ht]
\centerline{
\includegraphics[width=0.49\textwidth]{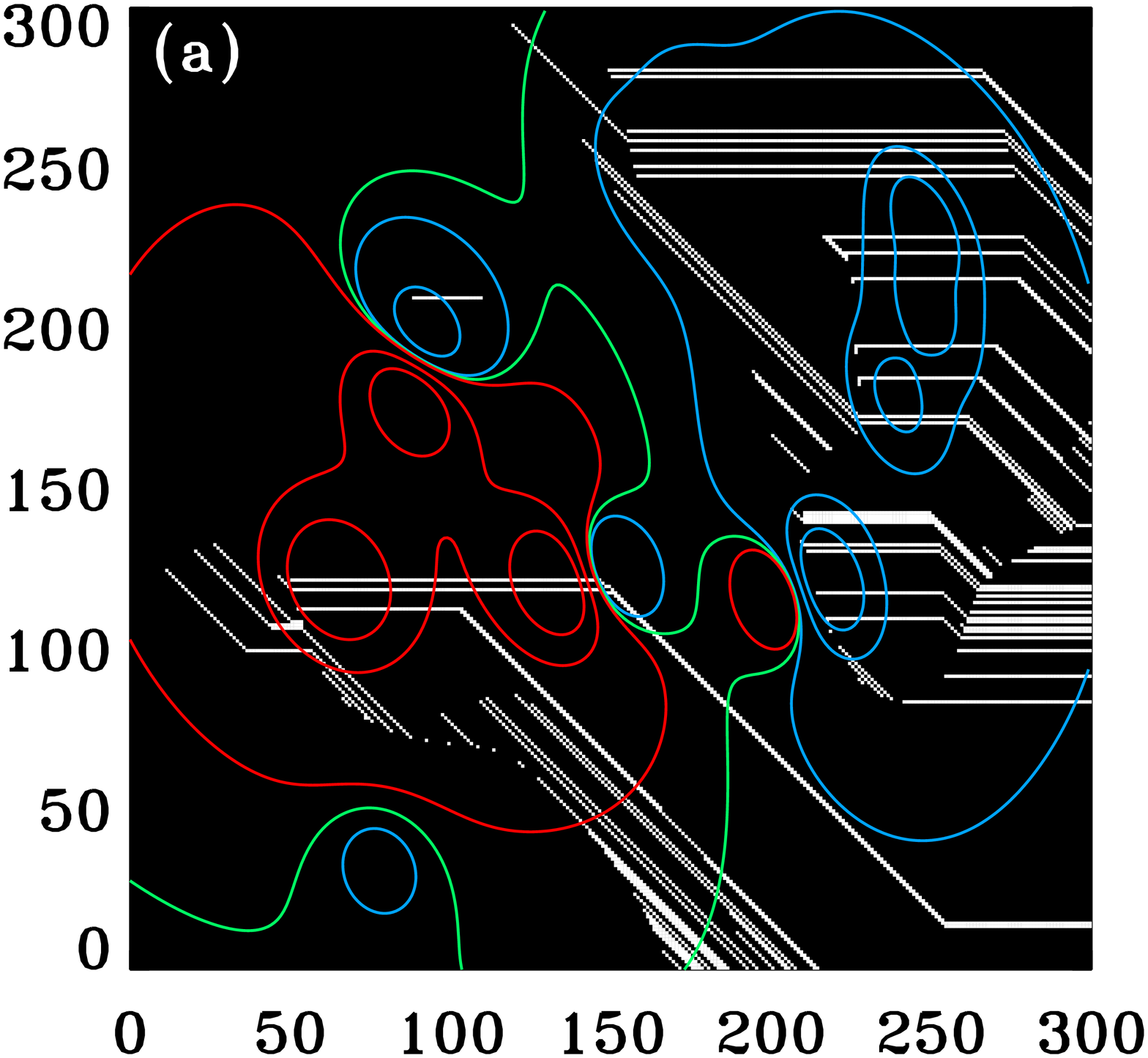}  
\hspace*{0.02\textwidth}
\includegraphics[width=0.49\textwidth]{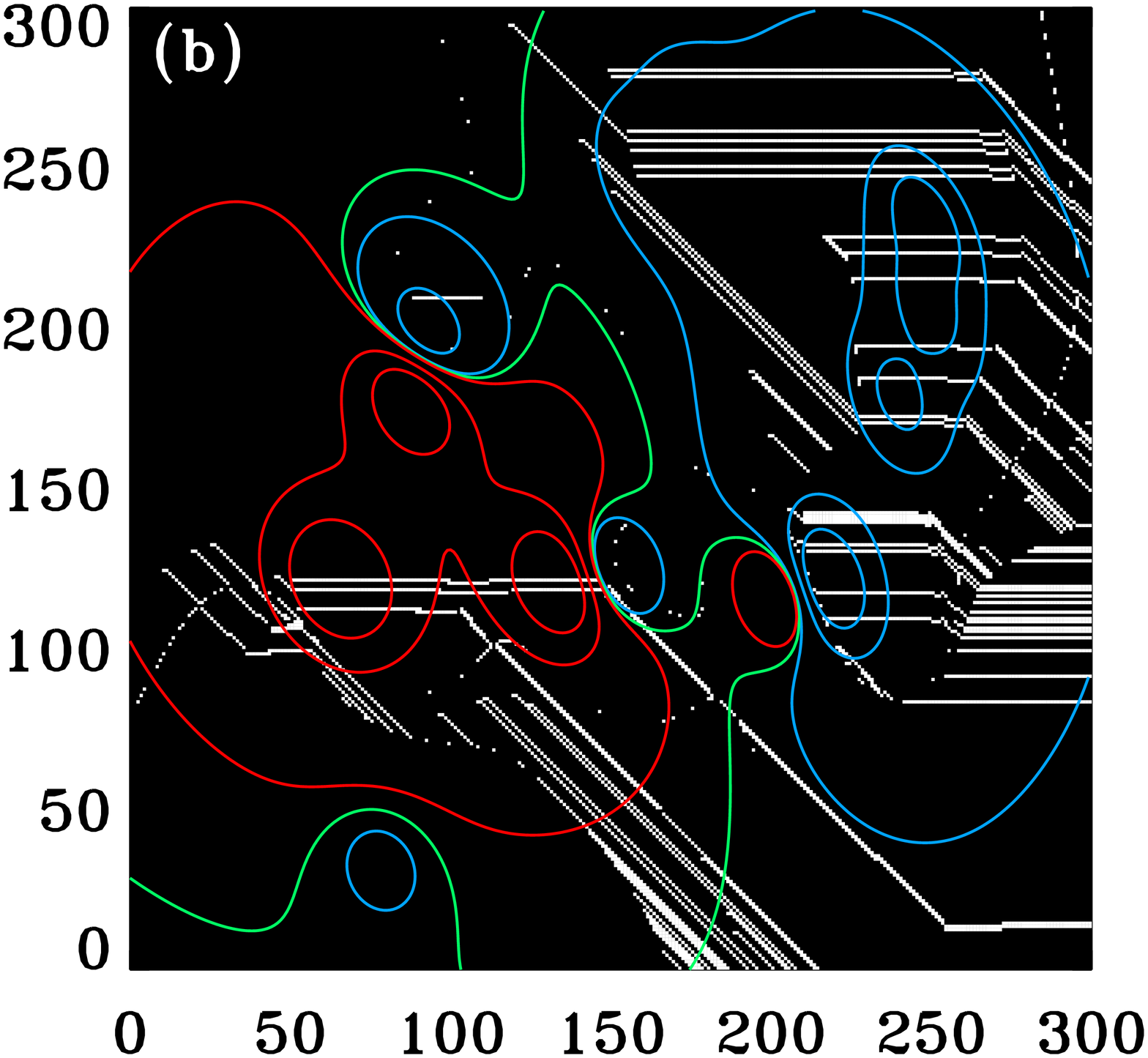}
}
\centerline{
\includegraphics[width=0.49\textwidth]{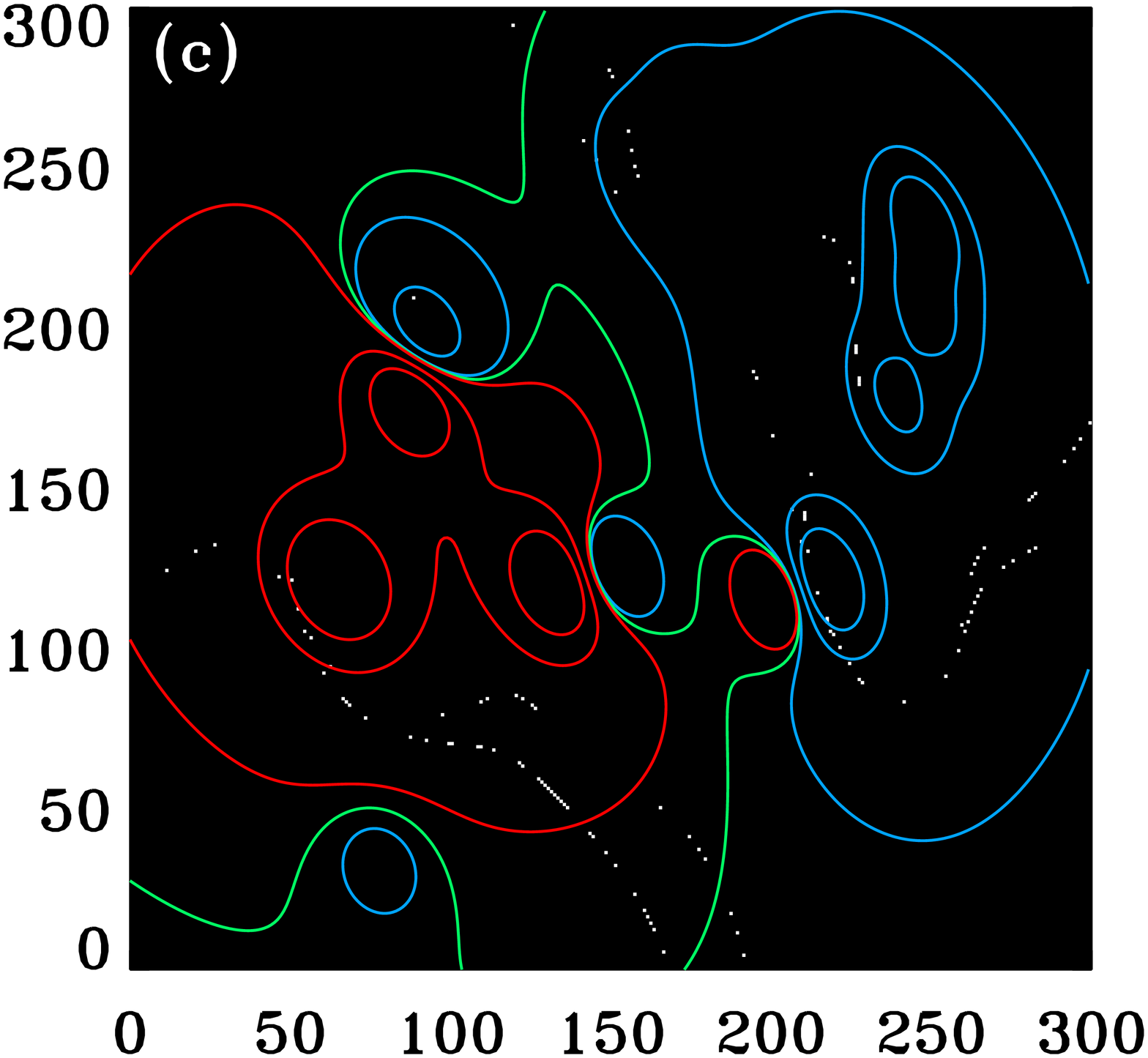}  
\hspace*{0.02\textwidth}
\includegraphics[width=0.49\textwidth]{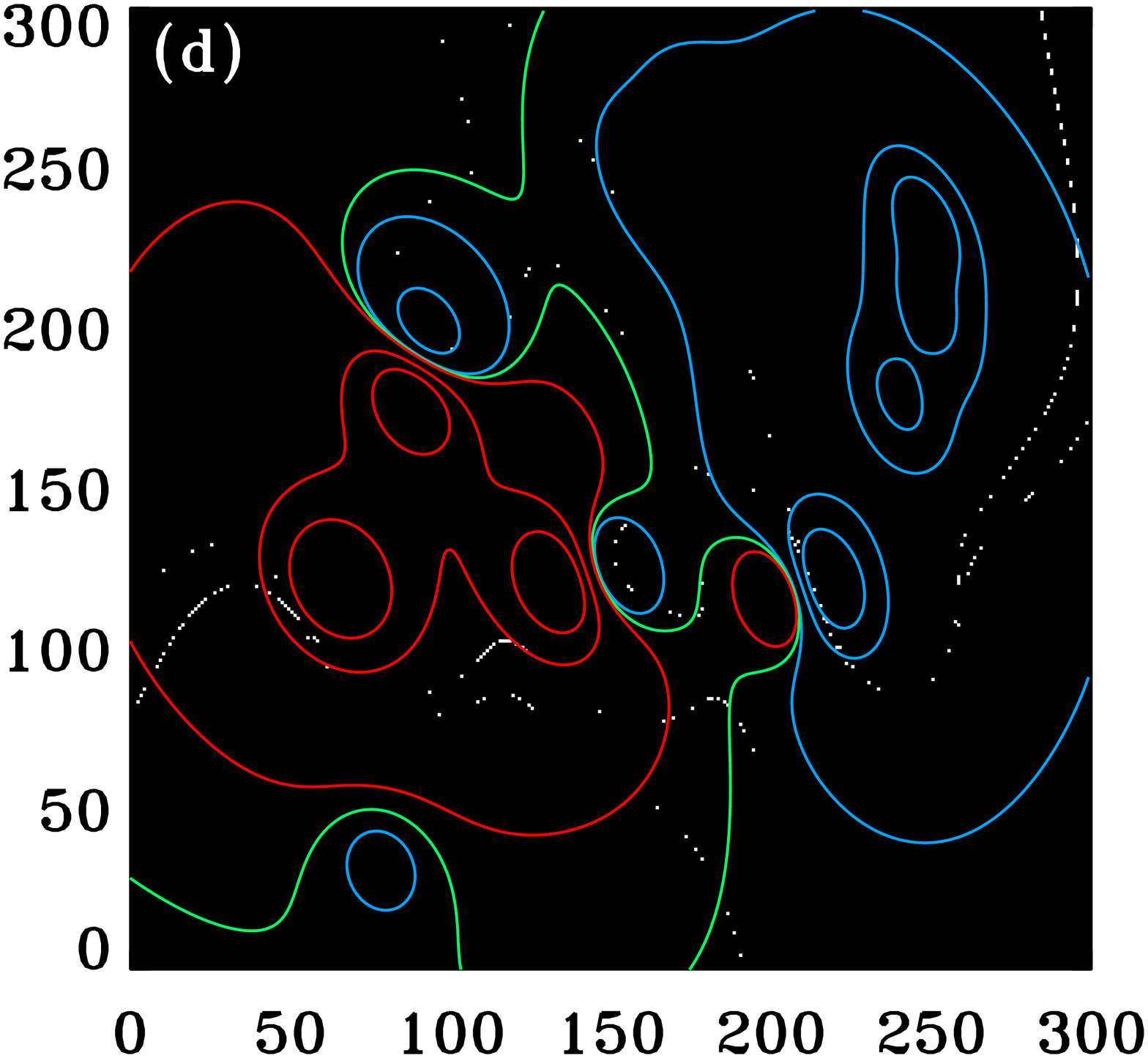}
}
\caption{(a) Disambiguation results for the sequential minimisation method applied to the  multipole field configuration, following Figure~\ref{compare_tpd7_wuai}(a).
(b) Same as (a) except for the second height, 1 pixel (or $0.5''$) above that shown in (a).
(c) Validity of the assumption made by the sequential minimisation method applied to the  multipole field, following Figure~\ref{compare_min_local}(b).
(d) Same as (c) except for the second height.
}
\label{compare_tpd7_min_local}
\end{figure}

The performance of this method is shown for the flux tube and arcade (Figure~\ref{compare_min_local}(a)). 
At disk centre the disambiguation at each height is independent because $\partial \bxi / \partial \zui$ and $\partial \byi / \partial \zui$ are not required to compute the divergence and the approximations for horizontal heliographic derivatives involve measurements at a single height; results for the lower height are displayed in Figure~\ref{compare_min_local}(a).
For the multipole field positioned away from disk centre (Figures~\ref{compare_tpd7_min_local}(a) and (b)) both heights are disambiguated simultaneously as required for off-disk-centre data.
Performance metrics for the lower height in each case are provided in Tables~\ref{fan_tab} and \ref{tpd7_tab}.

For each pixel the ambiguity is resolved by the stencil that references it first. 
Hence, for the pixels in a given stencil this algorithm assumes that the correct ambiguity resolution will be recovered for the unresolved pixels if the previously resolved pixels have the correct ambiguity resolution. 
We show areas for which this assumption is valid for both test cases in Figure~\ref{compare_min_local}(b) and Figures~\ref{compare_tpd7_min_local}(c) and (d).
Evidently, this assumption is not universally valid when the divergence is approximated from discrete observations.
However, the fraction of pixels that violate this assumption is smaller than the fraction that retrieve the incorrect azimuthal angle.
There are two reasons that this method fails at a particular pixel: 
{\it i)} the assumption is incorrect; and
{\it ii)} the propagation of an erroneous solution that originates in a region where the assumption is incorrect.
Propagation accounts for most of the areas with the incorrect azimuthal angle in Figure~\ref{compare_min_local}(a) and Figures~\ref{compare_tpd7_min_local}(a) and (b).
Most, but not all, pixels that violate the assumption initiate the propagation of  errors.
Also, there are pixels that violate the assumption but still retrieve the correct ambiguity resolution; these situations can occur when the assumption is violated and a neighbouring pixel has the incorrect ambiguity resolution, allowing the algorithm to get back onto the correct solution.

The algorithm described above is different from the implementation of \inlinecite{1999AA...347.1005B} in that they used four-point finite elements to approximate horizontal heliographic derivatives (we note that there may also be other differences).
We have tested other versions of this algorithm employing different approximations for the horizontal heliographic derivatives, such as four-point finite differences and finite elements. 
The results are very similar to those shown in Figure~\ref{compare_min_local} and \ref{compare_tpd7_min_local}: all versions examined violate the underlying assumption at some locations and subsequently propagate errors in the solution.

This method can propagate errors because it is sequential.
We have tried a non-sequential approach which assumes that $| ( \grad \vdot \B )_{i,j,k} |$ is minimised within the stencil based at each pixel regardless of the solution obtained in neighbouring pixels.
For both synthetic data sets the fraction of pixels for which this assumption is violated is actually larger than for the sequential approach.
Therefore, we conclude that the non-sequential approach does not substantially improve the performance of this method.

\subsection{The Global Minimisation Method}

In this section we present an iterative, global method which is based on the approximation for the divergence summed over the entire field of view.
This approach is related to the ``minimum energy'' algorithm for single-height data (see \myeg \opencite{metcalf94}; \opencite{metcalfetal06}; \opencite{lekaetal2008}; \opencite{hinode}).
We assume that the correct configuration of azimuthal angles over the field of view is the one that corresponds to the minimum of

\begin{equation}
E = \sum_{i=1}^{n_x}  \sum_{j=1}^{n_y} \left( | ( \grad \vdot \B )_{i,j,k} | +  | ( \grad \vdot \B )_{i,j,{k+1}} | \right) \, .
\label{esb}
\end{equation}

\noindent
Equation~(\ref{esb}) consists of two terms corresponding to the approximations for the divergence based at each of the two heights, $k$ and $k+1$.
For data positioned away from disk centre, where both heights are disambiguated simultaneously, we find that algorithms based on Equation~(\ref{esb}) consistently retrieve a better solution than algorithms based on a single term.
For the divergence at the first/second height the line-of-sight derivatives are approximated with forward/backward finite differences.
For each of these terms the horizontal heliographic derivatives are approximated using three-point finite differences as described in Section~\ref{sec_wuai}.
We have tested several different ways to approximate horizontal heliographic derivatives, such as four-point finite differences and three- and four-point finite elements, but these do not significantly affect the performance of the algorithm.

The number of possible solutions for this optimisation problem is finite but very large, $2^{n_x n_y}$, making a brute-force search for the global minimum of $E$ impractical.
One method that works well for large, discrete, global optimisation problems such as this is simulated annealing (\myeg \opencite{metrop}; \opencite{sa}; \opencite{nr}).
Using simulated annealing, the global minimisation method proceeds as follows.
A pixel is selected at random.
If changing the azimuthal angle at that pixel reduces the value of $E$ then the change is always accepted.
If changing the azimuthal angle increases the value of $E$ then the change is sometimes accepted, with a probability given by

\begin{equation}
p = \exp \left( - \Delta E / T \right) \, ,
\label{pup}
\end{equation}

\noindent
where $T$ is the ``temperature'' and $\Delta E$ is the change in $E$ (Equation~(\ref{esb})) caused by changing the azimuthal angle; positive values of $\Delta E$ imply that $E$ increases due to the reconfiguration.
The ability to sometimes accept an ``uphill'' reconfiguration enables simulated annealing to avoid getting stuck in a sub-optimal solution, provided that the cooling rate is sufficiently slow and the initial temperature is sufficiently high (\myeg \opencite{gg84}).

The annealing schedule (the initial temperature, the number of reconfigurations attempted at each temperature setting, and the cooling rate) can require some experimentation; in the interests of simplicity we use the following prescription.
The initial temperature, $T_0$, should greatly exceed any expected value of $\Delta E$.
Starting with the initial configuration of azimuthal angles we examine $100 n_x n_y$ reconfigurations of randomly chosen pixels (which are all accepted) and take two times the maximum value of $ | \Delta E |$ as the initial temperature.
It should be noted that this process effectively randomises the initial configuration of the azimuthal angles and therefore the initial configuration does not affect the final result (likewise, the high initial temperature further randomises the initial configuration because almost all reconfigurations will be accepted during early temperature increments).
At each temperature setting we examine $20 n_x n_y$ reconfigurations of the azimuthal angle at randomly chosen pixels.
The temperature varies according to $T_t = c^t T_0$, where $c$ is the cooling rate and $t$ is the label of the temperature increment; we use $c=0.999$.
We continue in this fashion until one of the following three conditions is satisfied:
{\it i)} no reconfigurations are accepted at the most recent temperature setting;
{\it ii)} the temperature is sufficiently low (less than $10^{-7} T_0$); or
{\it iii)} for ten consecutive temperature increments the value of $E$ does not change substantially (\myie $|E_t -E_{t-1}| / ( |E_t| + | E_{t-1}| ) < 10^{-5}$, where $E_t$ is the value of $E$ at the end of temperature increment $t$).

\begin{figure}[ht]
\centerline{
\includegraphics[width=0.49\textwidth]{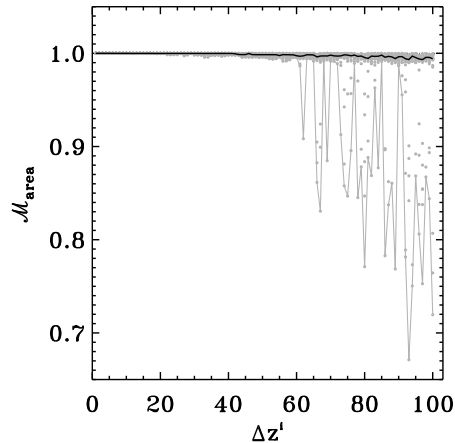}
}
\caption{The fraction of pixels correctly disambiguated at the lower height, $\mathcal{M}_{\rm area}$, as a function of vertical grid spacing $\Delta \zui$ (in units of horizontal pixels) for the global minimisation method with a fixed annealing schedule applied to the flux tube and arcade configuration (Figure~\ref{synth}(a)).
For each setting of $\Delta \zui$ twenty disambiguations are executed with different sequences of random numbers.
The {\it gray} dots represent the resultant $\mathcal{M}_{\rm area}$ value for each of these disambiguations.
The {\it black} line connects the median of the $\mathcal{M}_{\rm area}$ values for each $\Delta \zui$ setting and the {\it gray} lines connect the corresponding minimum and maximum of the $\mathcal{M}_{\rm area}$ values.
The {\it dotted} horizontal line at $\mathcal{M}_{\rm area}=1$ is included for reference.}
\label{dz_plot}
\end{figure}

For the flux tube and arcade positioned at disk centre, we show the fraction of pixels correctly disambiguated by the global minimisation method, $\mathcal{M}_{\rm area}$, as a function of the vertical grid spacing $\Delta \zui$ ( Figure~\ref{dz_plot}); only the lower height is evaluated because the disambiguations at each height are independent at disk centre for the derivative approximations employed here.
The simulation results of \inlinecite{2004ApJ...609.1123F} have a vertical grid spacing $\Delta \zui$ of $0.00625 L$ (equal to the horizontal grid spacing).
For the first height we use the snapshot displayed in Figure~\ref{synth}(a) at $\zui = 0.00625 L$. 
We allow the second height to vary, using all heights in the range $\zui = 0.0125 L, \ldots, 0.625 L$.
For each setting of $\Delta \zui$ we disambiguate twenty times with the algorithm described above but with a different sequence of random numbers for each disambiguation.

Independent disambiguations can behave differently because simulated annealing is a stochastic optimisation algorithm: the decision to move to higher $E$ is probabilistic (Equation~(\ref{pup})) and the order in which pixels are examined is random.
Consequently, if the annealing schedule is not ideal (\myeg if the cooling rate is too fast) the global minimum may not be found and a different answer may be obtained for a different sequence of random numbers.
To ensure that the global minimum is found for real magnetogram data it may be necessary to repeat the disambiguation process multiple times with different sequences of random numbers, a slower cooling rate, and/or with more attempted reconfigurations at each temperature setting.

For the flux tube and arcade at disk centre with $\Delta \zui = 1$~pixel the global minimisation method retrieves the correct azimuthal angle at every pixel for all twenty disambiguations with the imposed annealing schedule (see Table~\ref{fan_tab}).
Several points are worth noting regarding Figure~\ref{dz_plot}:
{\it i)} Broadly speaking, the performance decreases as $\Delta \zui$ increases.
{\it ii)} The scatter of the $\mathcal{M}_{\rm area}$ values increases as $\Delta \zui$ increases (most noticeably for $\Delta \zui \gtrsim 60$~pixels).
{\it iii)} For $\Delta \zui \gtrsim 3$~pixels the correct solution is no longer found for all twenty disambiguations.
The maximum of the $\mathcal{M}_{\rm area}$ values is one at this point so the correct solution could be obtained simply by disambiguating multiple times and selecting the solution with the smallest value of $E$.
{\it iv)} The point where the maximum of the $\mathcal{M}_{\rm area}$ values deviates from one is $\Delta \zui \approx 10$~pixels. 
Beyond this point, a perfectly correct solution is not obtained for any of the twenty disambiguations with the imposed annealing schedule, yet for the largest values of $\Delta \zui$ the median value for $\mathcal{M}_{\rm area}$ is still greater than 0.99.

\begin{figure}[ht]
\centerline{
\includegraphics[width=0.49\textwidth]{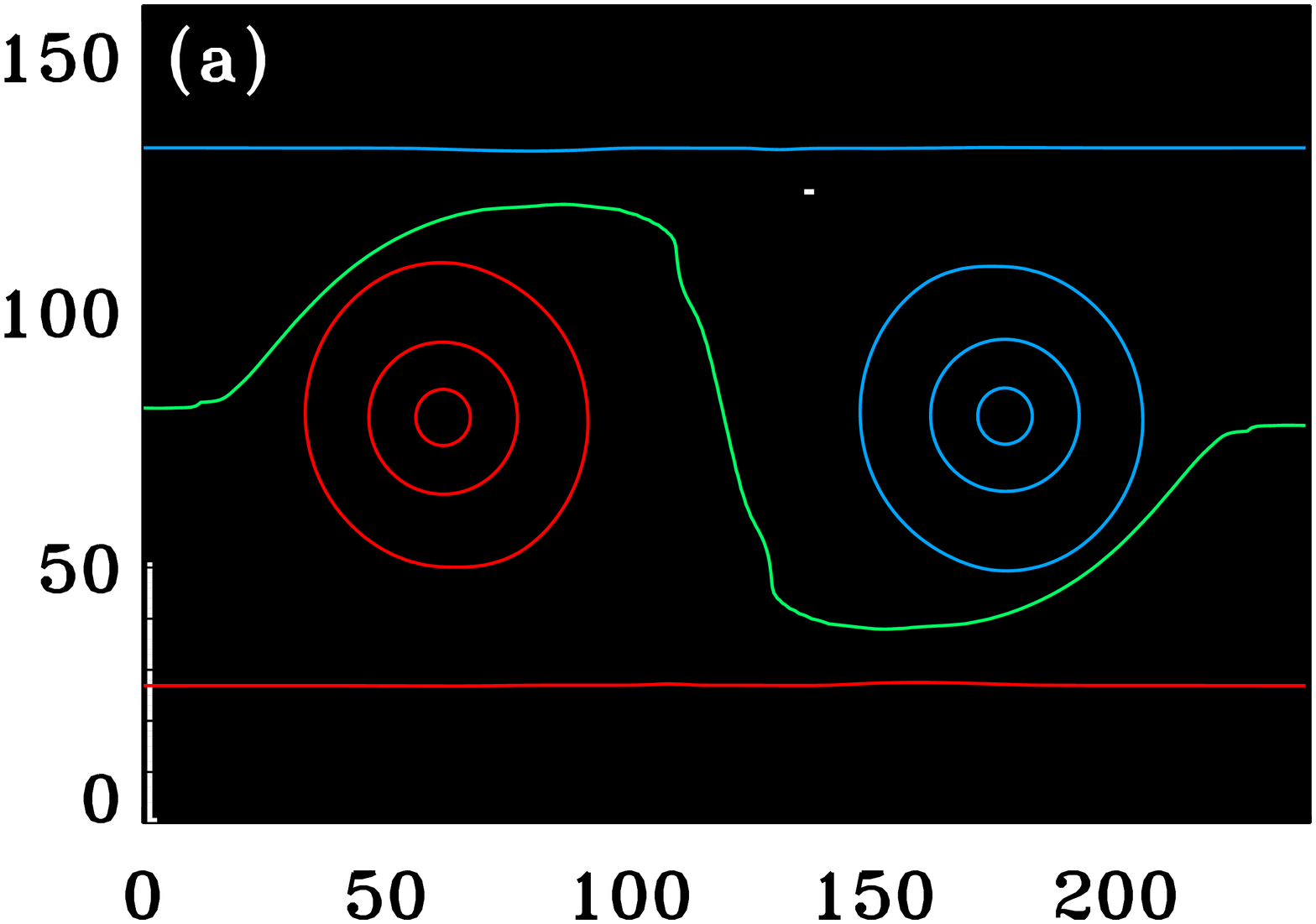}  
\hspace*{0.02\textwidth}
\includegraphics[width=0.49\textwidth]{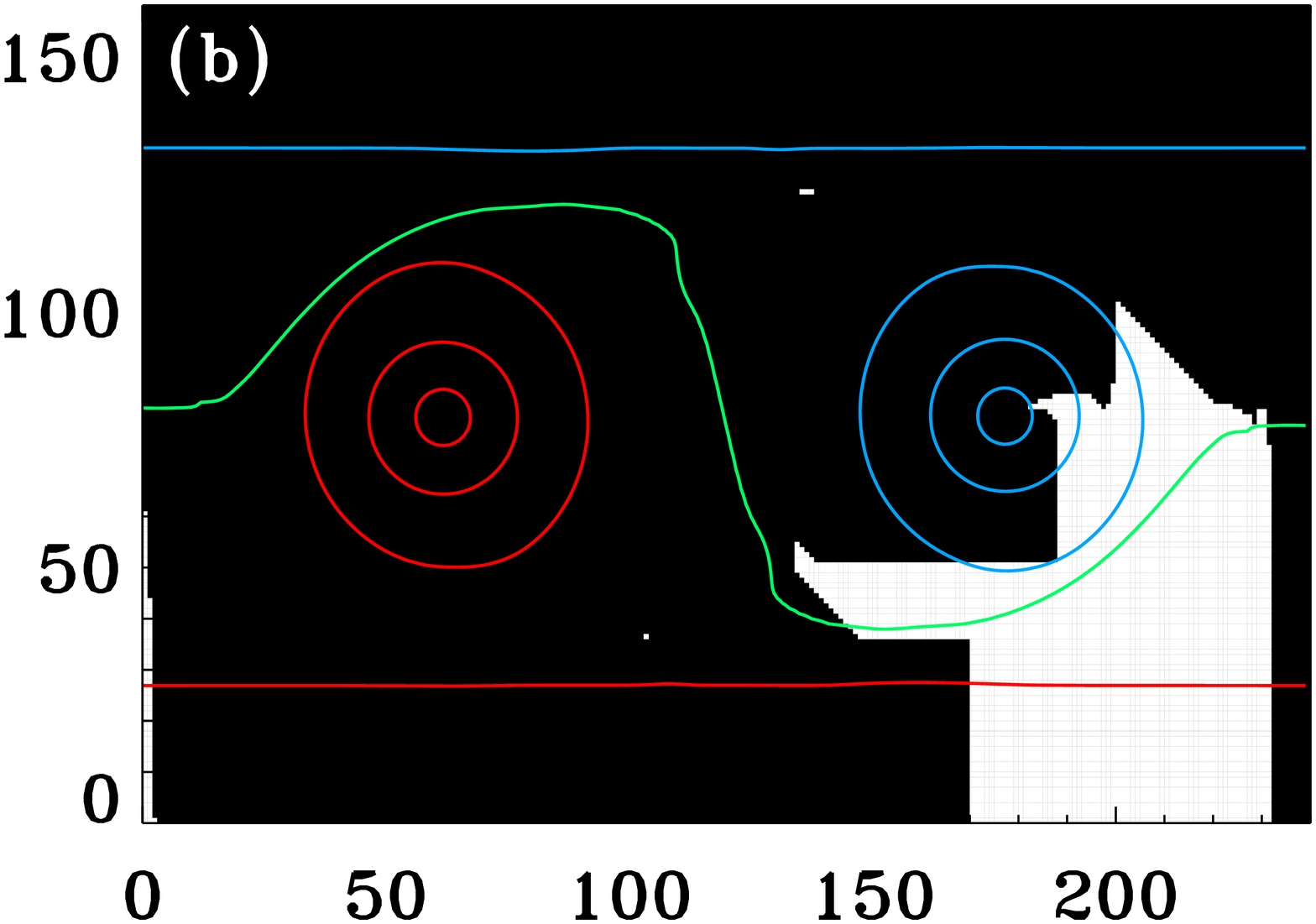} 
}
\caption{Disambiguation results for the global minimisation method  applied to the flux tube and arcade configuration with $\Delta \zui = 65$~pixels, following Figure~\ref{fan_ts56}.
(a) The solution retrieved with the best $\mathcal{M}_{\rm area}$ value.
(b) The solution retrieved with the worst $\mathcal{M}_{\rm area}$ value.
}
\label{compare_fan_me}
\end{figure}

For large values of $\Delta \zui$ the difference between the best and worst solutions retrieved by the global minimisation method can be significant (Figure~\ref{compare_fan_me}).
The main cause of such deviations is the annealing schedule.
We have conducted experiments that show that more consistent and better results can be obtained at large values of $\Delta \zui$ with annealing schedules that involve slower cooling rates or more attempted reconfigurations at each temperature setting.

\begin{figure}[ht]
\centerline{
\includegraphics[width=0.49\textwidth]{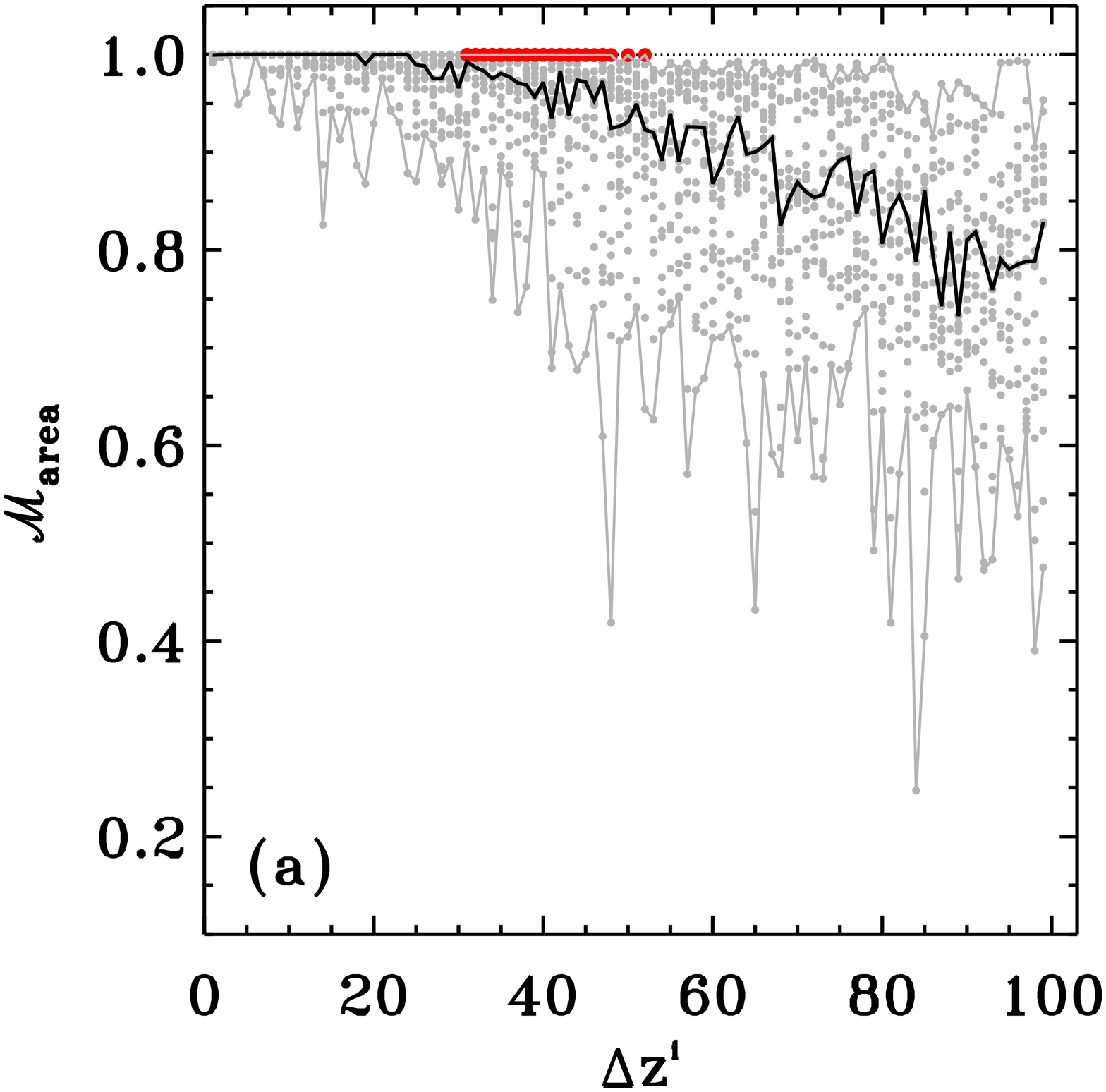}  
\hspace*{0.02\textwidth}
\includegraphics[width=0.49\textwidth]{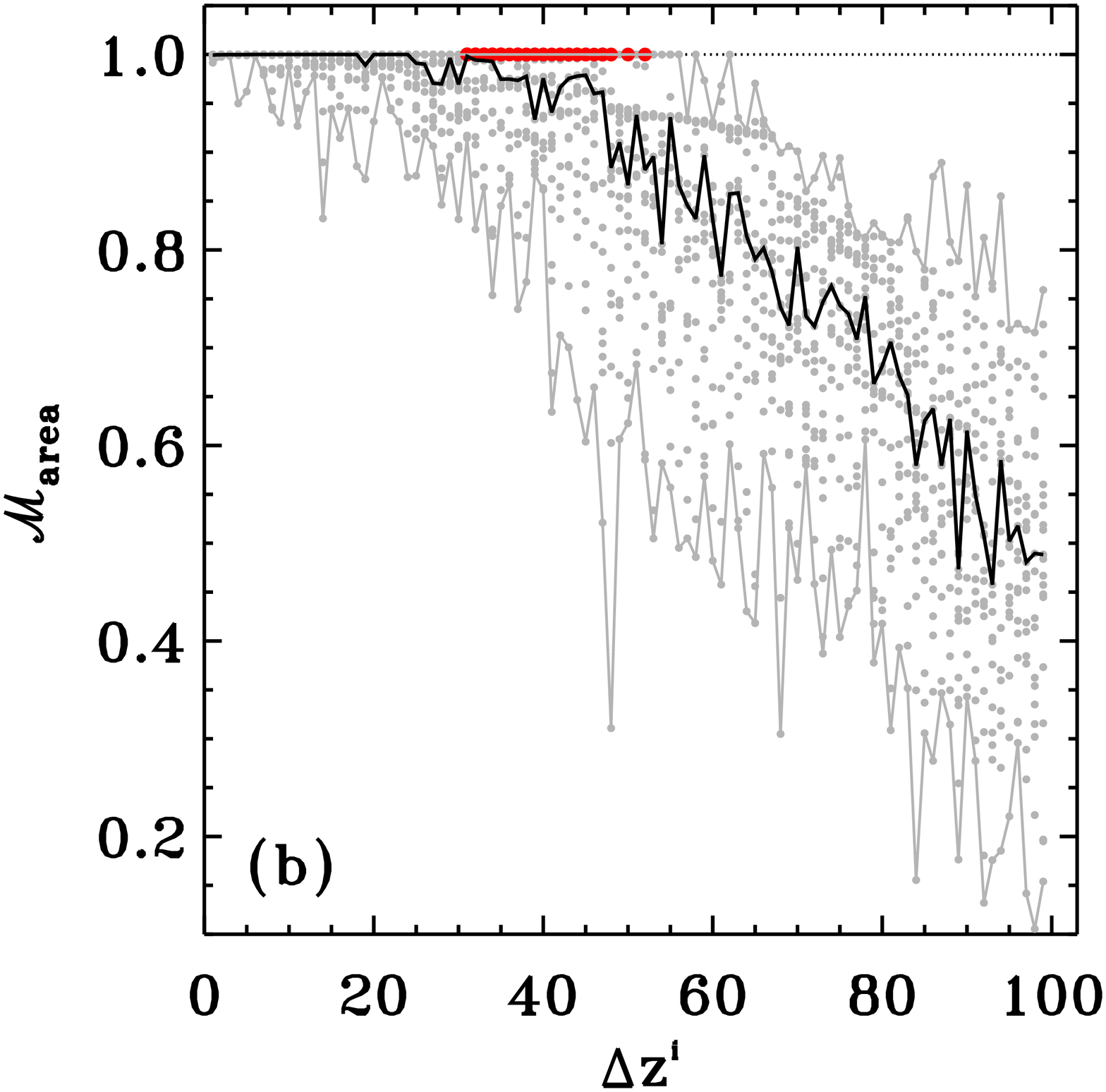}
}
\caption{(a) The fraction of pixels correctly disambiguated at the lower height, $\mathcal{M}_{\rm area}$, as a function of vertical grid spacing $\Delta \zui$ (in units of horizontal pixels) for the global minimisation method with a fixed annealing schedule applied to the multipole field configuration in Figure~\ref{synth}(b), following Figure~\ref{dz_plot}.
The {\it red} dots represent the $\mathcal{M}_{\rm area}$ values for each of the disambiguations that retrieved a solution with $E$ less than that for the correct solution.
(b) Same as (a) except for the second height.
}
\label{dz_plot2}
\end{figure}

For the multipole field positioned away from disk centre the variation of $\mathcal{M}_{\rm area}$ as a function of the line-of-sight grid spacing $\Delta \zui$ is shown (Figure~\ref{dz_plot2}); both heights are disambiguated simultaneously as required for off-disk-centre data.
For the first height we use the field displayed in Figure~\ref{synth}(b).
We allow the second height to vary in increments along the line-of-sight direction corresponding to the horizontal pixel size, $0.5''$.
Of the twenty  disambiguations with  $\Delta \zui =1$~pixel, only two solutions obtained $\mathcal{M}_{\rm area}=1$ for both heights (see Table~\ref{tpd7_tab}); the remainder of solutions obtained $\mathcal{M}_{\rm area}=0.99$ or better at each height.

The performance of the global minimisation method applied to the multipole field configuration (Figure~\ref{dz_plot2}) degrades more rapidly as $\Delta \zui$ increases than it does for the flux tube and arcade (Figure~\ref{dz_plot}).
This is partly because the optimisation problem for data positioned away from disk centre is substantially more challenging, as both heights must be disambiguated simultaneously.
Differences in the variation of the two fields with height may also contribute to the contrasting performance.

The synthetic data for the multipole field configuration include the effects of curvature but we have not included curvature in the treatment of the divergence (Equations~(\ref{divb2}) and (\ref{da})). Experiments with two synthetic data sets that have the same field configuration but include and exclude curvature indicate that curvature does not significantly influence the performance of the global minimisation algorithm, provided that the field of view is much smaller than the radius of curvature of the Sun.

\begin{figure}[ht]
\centerline{
\includegraphics[width=0.49\textwidth]{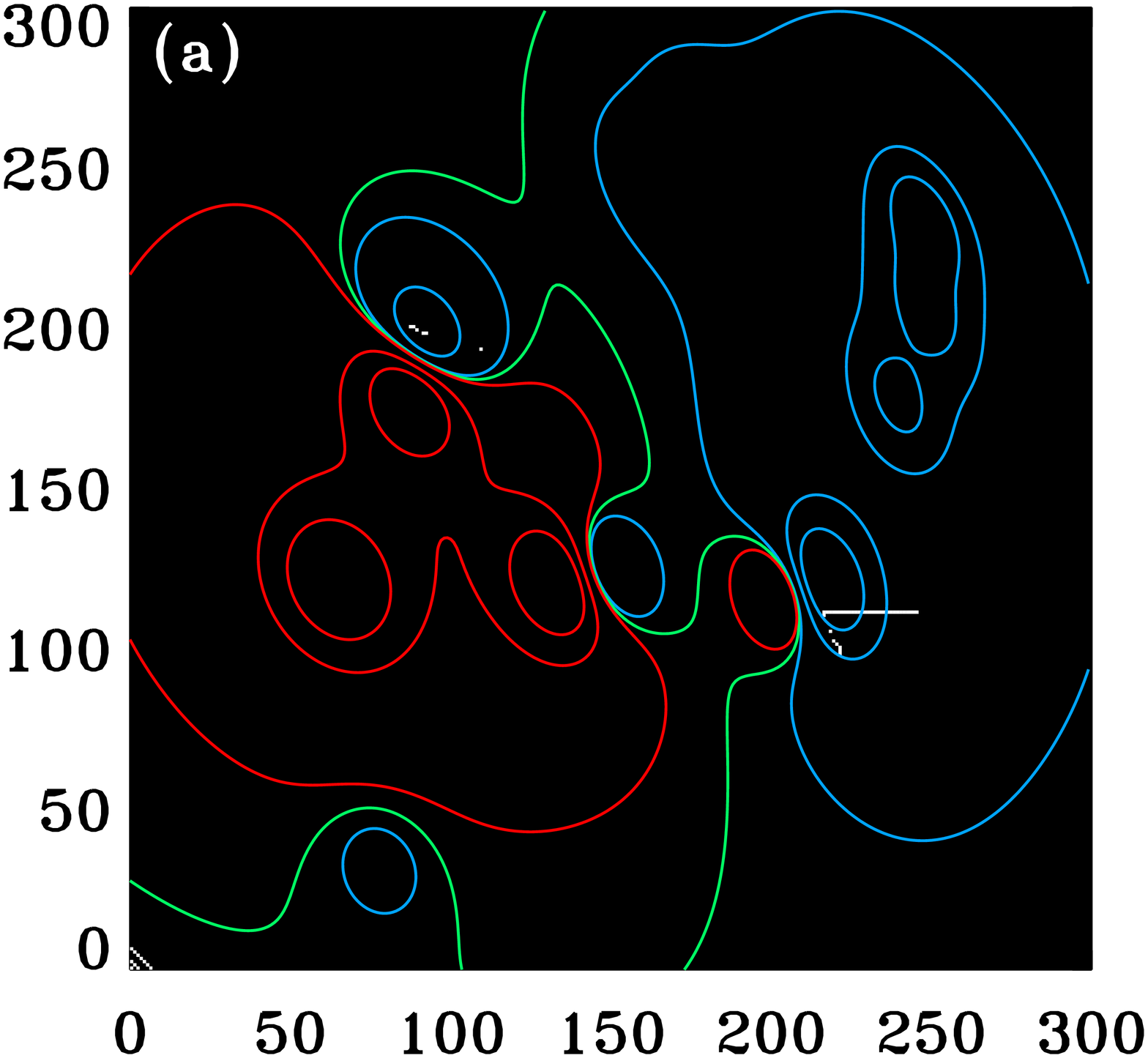}  
\hspace*{0.02\textwidth}
\includegraphics[width=0.49\textwidth]{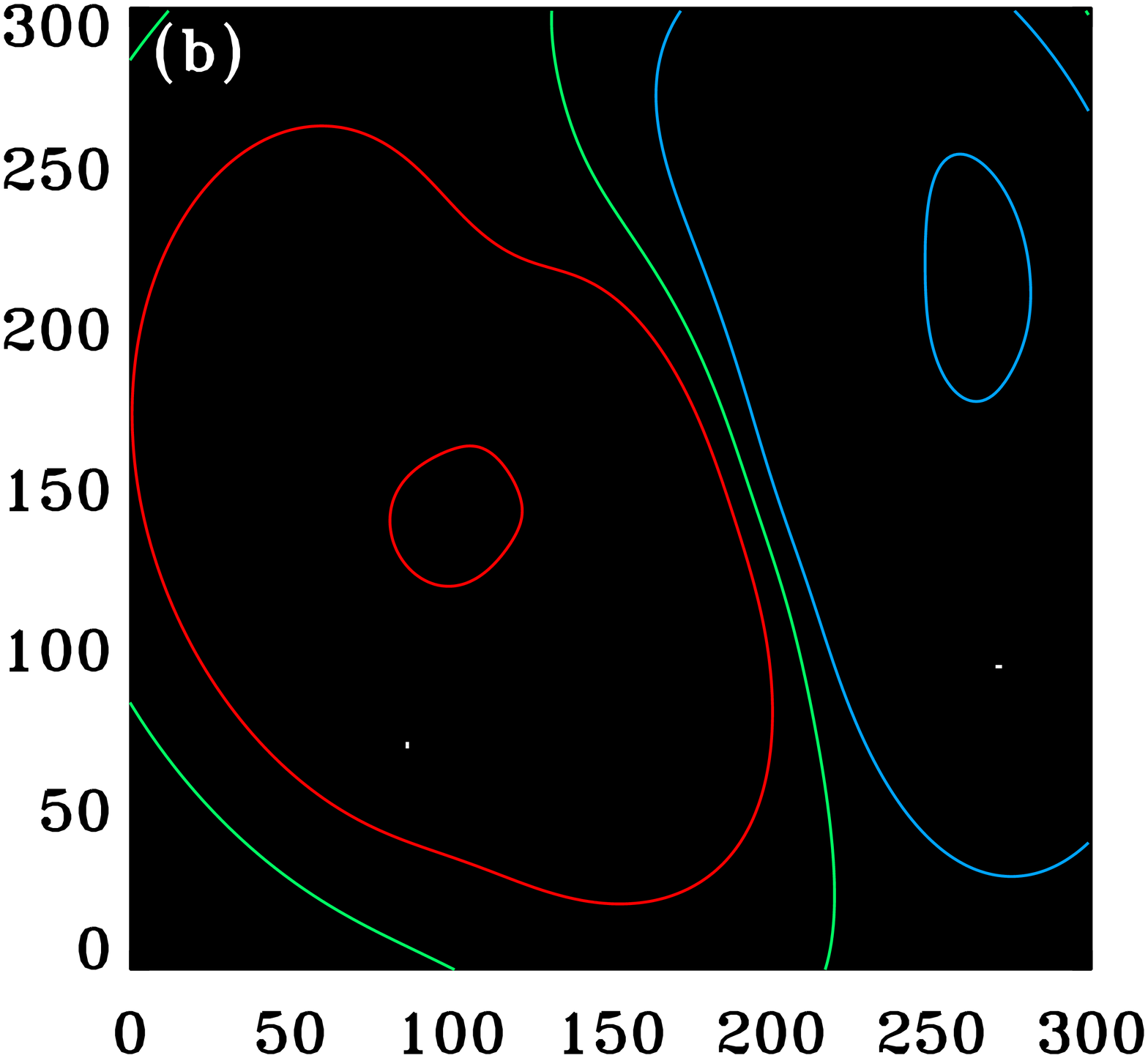}
}
\caption{(a) Disambiguation results for the global minimisation method applied to the multipole field configuration with $\Delta \zui =50$~pixels, following Figure~\ref{compare_tpd7_wuai}(a).
For this particular case $E$ for the retrieved solution is less than that for the correct solution.
(b) Same as (a) except for the second height, 50 pixels above that shown in (a).
}
\label{compare_tpd7_2}
\end{figure}

In Figure~\ref{dz_plot2}, we see that for some values of $\Delta \zui$ larger than 31~pixels ({\it red} dots in Figure~\ref{dz_plot2}), the algorithm retrieves solutions with $E$ less than that for the correct solution.
Thus, the assumption that the correct solution corresponds to the configuration of azimuthal angles that minimises $E$ (Equation~(\ref{esb})) is violated in these cases.
Nevertheless, the performance of the resulting solutions is quite good, with  $\mathcal{M}_{\rm area}$ values greater than 0.99 at both heights (see \myeg Figure~\ref{compare_tpd7_2}).
For $\Delta \zui$ larger than 53~pixels no such solutions are found, which  probably indicates that the fixed annealing schedule employed here is less than ideal for large values of $\Delta \zui$.

Results for a case where the retrieved solution has $E$ less than that for the correct solution are shown for the multipole configuration with $\Delta \zui =50$~pixels (Figure~\ref{compare_tpd7_2}).
Interestingly areas that fail occur at different locations at each height; this situation is fairly typical for $\Delta \zui  \gtrsim 25$~pixels.
For smaller values of $\Delta \zui$ incorrect regions tend to occur at similar locations at both heights.

For the flux tube and arcade  we have not found a solution for which $E$ is less than that for the correct solution, in the case where the horizontal resolution is fixed but the vertical resolution varies (Figure~\ref{dz_plot}).
We find that this situation does arise for the flux tube and arcade when both the horizontal and vertical sampling are about a factor of seven times coarser than on the original grid.
Clearly, the validity of the underlying assumption made by the global minimisation method -- that the correct configuration of azimuthal angles corresponds to the minimum value of $E$  -- is sensitive to the spatial resolution of the observations. 
This suggests that for real magnetogram data (that may contain noise and/or unresolved structure) additional constraints or criteria may be required to determine if the global minimum corresponds to the correct solution. 
It may also be possible to use measurement error information to determine the uncertainties associated with the configuration corresponding to the global minimum.

\section{Conclusions}
\label{sec_conc}

We investigate how the divergence-free property of magnetic fields can be exploited to resolve the azimuthal ambiguity that is present in solar vector magnetogram data by using line-of-sight and horizontal heliographic derivative information approximated from discrete measurements.
Using synthetic data we test several methods that each make different assumptions about how to use the divergence to resolve the ambiguity.
We find that all of the assumptions considered can be violated when the divergence is approximated from discrete measurements (depending on the nature of the field and the resolution of the observations), but some assumptions are more robust than others.

The \citeauthor{wuai1990} criterion expresses the divergence-free condition as an inequality (\myeg \opencite{wuai1990}; \opencite{1993AA...278..279C}; \opencite{1993AA...279..214L}; \opencite{lietal07}).
We find that methods based on this criterion are very sensitive to the smoothness of the initial configuration of azimuthal angles and the order in which pixels are examined.
In addition, the underlying assumption can be incorrect over a significant fraction of the field of view when derivatives are approximated from discrete measurements.
The sequential minimisation method (\opencite{1999AA...347.1005B}) assumes that the magnitude of the divergence is minimised over the pixels used to approximate the divergence at each point.
Our tests indicate that this approach is more robust than the \citeauthor{wuai1990} criterion but can produce errors when the underlying assumption is violated.
Moreover, the algorithm can propagate erroneous solutions into regions that do not violate the underlying assumption.

We find that the most promising algorithm is the global minimisation method which assumes that the correct configuration of azimuthal angles corresponds to the minimum of the magnitude of the divergence summed over the entire field of view.
This method relies on the  magnitude of the divergence.
Therefore, with this method, the {\it sign and magnitude} of the line-of-sight derivatives of all three components of the magnetic field vector are required to resolve the azimuthal ambiguity away from disk centre.

For the synthetic data employed here, the field is sampled at discrete spatial locations where the exact value of the field is provided. Thus, the effects of noise and unresolved structure in both the horizontal heliographic and line-of-sight directions are neglected. The performance of the global minimisation method is very promising, but before this approach can be reliably applied to real magnetogram data its performance should be tested with more realistic synthetic data that include the effects of noise and unresolved structure (\myeg \opencite{lekaetal2008}).

\begin{acks}
The authors would like to thank Yuhong Fan for providing one of the synthetic data sets used in this investigation.
This work was supported by funding from NASA under contracts NNH05CC75C and NNH09CE60C.
\end{acks}


\end{article}
\end{document}